\documentclass[11pt,a4paper]{article}
\usepackage[margin=24mm]{geometry}
\usepackage[T1]{fontenc}
\usepackage{lmodern,graphicx,amsmath,amssymb,booktabs,array,xcolor,microtype}
\usepackage[numbers,sort&compress]{natbib}
\usepackage[colorlinks=true,allcolors=blue!45!black]{hyperref}
\usepackage[font=small,labelfont=bf]{caption}
\newcommand{\wn}{\ensuremath{\mathrm{cm}^{-1}}}
\newcommand{\Ntwo}{\ensuremath{\mathrm{N}_2}}
\newcommand{\rtwo}{r$^2$SCAN}

\title{\Large\bfseries Neutron Spectroscopy Transformed: Deuterated ZIF-8 and Machine-Learned Potentials Catch Nitrogen Threading Its Gates}
\author{Jeff Armstrong$^{1,2,}$\thanks{Corresponding author: \href{mailto:jeff.armstrong@stfc.ac.uk}{jeff.armstrong@stfc.ac.uk}}, Hamish Cavaye$^{1}$, Pankaj Sharma$^{3}$ and Matthew E. Potter$^{2}$\\[6pt]
\parbox{0.92\textwidth}{\centering\small
$^{1}$ISIS Neutron and Muon Source, STFC, Rutherford Appleton Laboratory, Harwell Campus, Didcot OX11 0QX, United Kingdom\\
$^{2}$Department of Chemistry, University of Bath, Bath BA2 7AY, United Kingdom\\
$^{3}$United Kingdom Atomic Energy Authority, Culham Campus, Abingdon, Oxfordshire OX14 3DB, United Kingdom}}
\date{}

\usepackage{longtable,colortbl}
\usepackage{tocloft}
\newcommand{\sifigure}[4][.98]{\begin{figure}[htbp]\centering\includegraphics[width=#1\linewidth,height=.70\textheight,keepaspectratio]{#2}\caption{#3}\label{#4}\end{figure}}
\definecolor{atomc}{HTML}{15803D}\definecolor{cartc}{HTML}{0891B2}\definecolor{fwc}{HTML}{1D4ED8}
\definecolor{gc}{HTML}{9333EA}\definecolor{modec}{HTML}{BE123C}\definecolor{kc}{HTML}{C2410C}
\definecolor{fwfill}{HTML}{DCE6F5}\definecolor{gfill}{HTML}{E6D9F5}\definecolor{cfill}{HTML}{F3EEFA}\definecolor{kfill}{HTML}{F8D9C6}

\begin{document}
\maketitle

\begin{abstract}
We report a step change in what inelastic neutron scattering (INS) can reveal, by combining machine-learned interatomic potentials (MLIPs) with deliberate deuteration of the sample. INS's prowess in charting framework and guest motions within porous materials during in situ dosing has historically had one caveat: costly density-functional-theory calculations are a major bottleneck in the speed and depth of analysis. We show that rapid frontier MLIPs remove this bottleneck on both counts, keeping pace with the experiment while bringing a new level of insight. Deuterating the linkers of ZIF-8 pries apart bands that overlap in the hydrogenous framework, and we test four MLIPs against eight stages of nitrogen dosing at 77~K. On loading, a previously unobserved band appears at 270~\wn\ and grows gradually through the dosing series. The simulations trace it to single \Ntwo\ molecules threaded end-on through four-ring windows, each turning its ring of linkers and propping the window open, and unbiased dynamics catch molecules threading into the windows on their own, finding a free-energy trap that crowding stabilizes. This is the first vibrational signature of the threaded state and, crucially, its intensity directly counts the threading over the slow, macroscopic timescale of gas uptake, in situ. More broadly, the simulations explain how each framework band responds to its guests, establishing MLIP-assisted inelastic neutron scattering as a route to mechanism, not just assignment.
\end{abstract}

\medskip
Inelastic neutron scattering (INS) can give a detailed, mode-by-mode account of the vibrations of a porous material and its guests as the material is progressively loaded with guest molecules, from small gases to larger organic molecules.\cite{Easun2017,Morton2025,OMalley2018} It has no selection rules: every mode appears, with an intensity set by the motion of the atoms alone, so a spectrum can be calculated directly from the vibrations. Neutrons penetrate deeply, through the sample cell and into the bulk of a powder, so a framework can be dosed and measured in situ; and the technique is especially sensitive to hydrogen, so organic and hybrid materials give strong spectra quickly. Historically, INS has relied on density-functional theory (DFT) for interpretation, and DFT has relied on INS for validation. The weak link is time: DFT lattice dynamics of a single guest-loaded cell of several hundred atoms costs days to weeks of supercomputer time, so the analysis typically lags well behind the measurement. Treating several gas loadings systematically is prohibitively expensive, all the more because each loading needs many guest configurations to be explored before a global minimum is found, or even confidently approached; and molecular dynamics over the relevant timescales is effectively out of reach. Machine-learned interatomic potentials (MLIPs) change this completely.\cite{MACE,MACEMH,UMA,OMat24,OMol25,MOFPhonons2025,uMOF} They deliver forces close to DFT orders of magnitude faster: the same guest-loaded cell is relaxed and its full force-constant matrix computed in about an hour on a single GPU, and nanoseconds of molecular dynamics on cells of this size, far beyond the reach of DFT, become possible. The far harder test is the curvature of the potential-energy surface, for the host, the guests and the couplings between them, which an INS spectrum probes directly and which matching energies and forces to DFT does not guarantee.\cite{MOFSimBench,Armstrong2022} If MLIPs have truly reached this level of accuracy, the calculation is no longer the slow step, and simulation becomes an active partner in the experiment, helping to steer it while it is still running.

A second shift in paradigm is the sample itself. Vibrational neutron spectroscopy has stuck stubbornly to hydrogenous samples, because the large cross section of hydrogen (82~barns, against 7.6 for deuterium\cite{NeutronXS}) is what makes the signal strong; deuteration, routine in neutron diffraction, is rarely exploited. Yet replacing hydrogen by deuterium does far more than weaken a spectrum. Because framework modes are heavily coupled, they do not simply scale with the square root of the mass: vibrational character is redistributed among them and peaks shift in non-trivial ways, which calculation predicts, as we showed recently for ZIF-8.\cite{Isotope2026} This is exactly what makes deuteration powerful. Two modes that overlap in the hydrogenous material but are coupled differently to the hydrogen atoms shift by different amounts on deuteration, and are pried apart. Used deliberately, deuteration turns from a loss of signal into a tool for resolving bands that would otherwise stay hidden.\cite{Armstrong2021}

ZIF-8 is an ideal test of both. Its 2-methylimidazolate linkers swing open the windows between cages, under pressure and on nitrogen adsorption at 77~K, giving a stepped uptake whose molecular origin has long been debated.\cite{Moggach2009,Fairen2011,Zhang2013,Ania2012,Tian2016,Coudert2017} Simulations\cite{Fairen2011,FairenJimenez2012} and diffraction\cite{Tian2016,Hobday2018} place guests in and beside the small four-membered-ring windows at high loading, methyl-rotor INS suggests that guests there stop the linkers turning back,\cite{Titov2023} and in situ INS has followed gate opening under nitrogen and carbon dioxide.\cite{Mor2025,Bahadur2025} What a molecule in such a window does to the framework, and how it shows in a vibrational spectrum, is unknown: in hydrogenous ZIF-8 the mid-energy framework bands overlap, so INS under nitrogen shows them changing but cannot say how or why.\cite{Casco2016,Mor2025} Here we follow deuterated ZIF-8 through a nitrogen dosing series, show that four frontier MLIPs, none fitted to these data, reproduce its spectrum band for band without frequency scaling, and use them to trace a new band to single nitrogen molecules, each threaded through a window and holding a ring of linkers open. The band, hidden in hydrogenous ZIF-8, is the first vibrational signature of the window site found by diffraction, and the peak's intensity is a direct measure of how many nitrogen molecules have been threaded into the four-membered rings. Unbiased dynamics then show molecules threading the windows on their own, through a gate guarded by two linkers, into a free-energy trap, and the slow, window-by-window capture explains why the uptake itself slows as the band grows. A single guest prising a ring of linkers open thus becomes something a spectrum can measure. The agreement is itself a stringent validation of frontier MLIPs: because every mode appears in INS, with an intensity set by the atomic motion alone, the spectrum tests the full force-constant matrix of host, guests and their coupling, a test that energies, forces or optical spectra with their selection rules cannot provide.

\begin{figure}[tbp]\centering
\includegraphics[width=\linewidth]{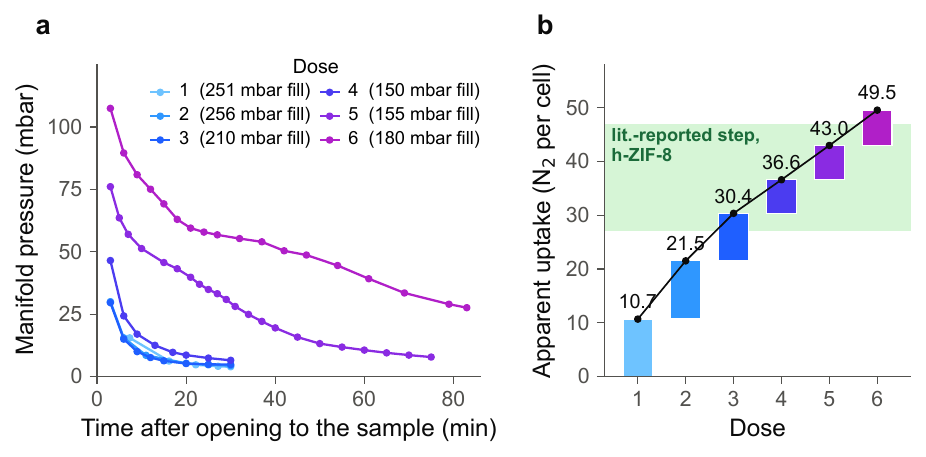}\\[2pt]
\includegraphics[width=\linewidth]{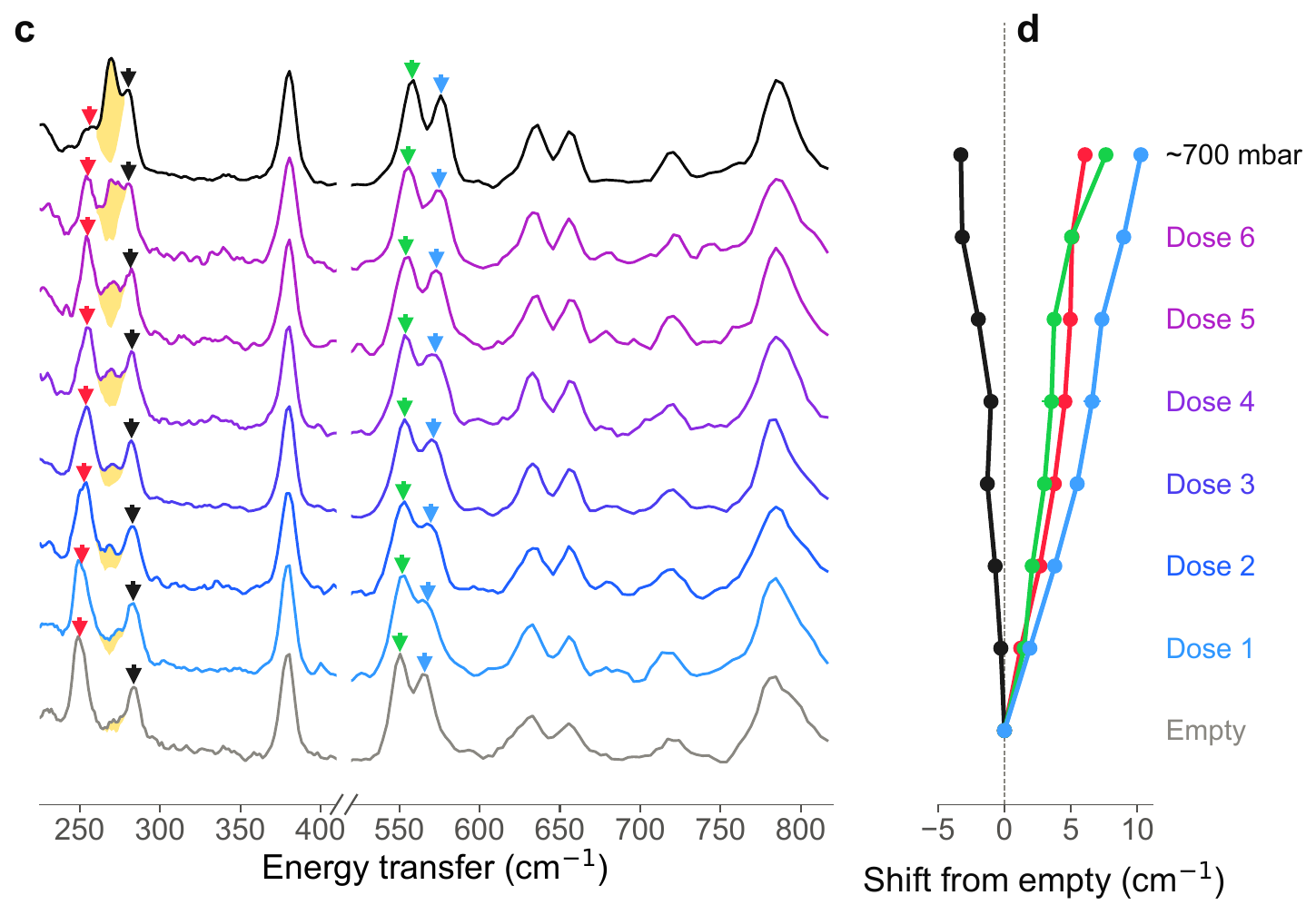}
\caption{Dosing nitrogen into deuterated ZIF-8 at 77~K. (a) Manifold pressure during each of six doses. (b) Apparent uptake per dose (bars) and cumulative uptake (line); green, the uptake-step range reported for hydrogenous ZIF-8.\cite{Tian2016} (c) INS spectra from the empty framework to about 700~mbar; triangles, fitted bands; yellow, the new band near 270~\wn. (d) Shift of each band from the empty framework.}
\label{fig:exp}
\end{figure}

Nitrogen was dosed in six steps into fully deuterated ZIF-8 at 77~K, and INS was measured on TOSCA\cite{Parker2014,Pinna2018,TOSCAinstrument} after each dose, spanning apparent loadings of 11 to 50 \Ntwo\ per cell across the reported uptake step\cite{Tian2016} (Figure~\ref{fig:exp}a,b; SI Section~\ref{SI-si:dosing}). The spectrum responds at every dose (Figure~\ref{fig:exp}c,d). Three linker bands, near 250, 550 and 566~\wn, stiffen as the guests crowd in, much as one would expect. The fourth, the Zn--N stretch at 283~\wn, does the opposite: it softens steadily as the loading rises. The framework is not simply being squeezed stiffer; it is reorganizing. Most strikingly, a new band grows at about 270~\wn, in the valley between them, until it carries half of the intensity in this region. The uptake itself changes character as the framework fills: the pressure decays exponentially for the first four doses but not for the last two (Figure~\ref{fig:exp}a), a clue we return to below.

\begin{figure}[tbp]\centering
\includegraphics[width=\linewidth]{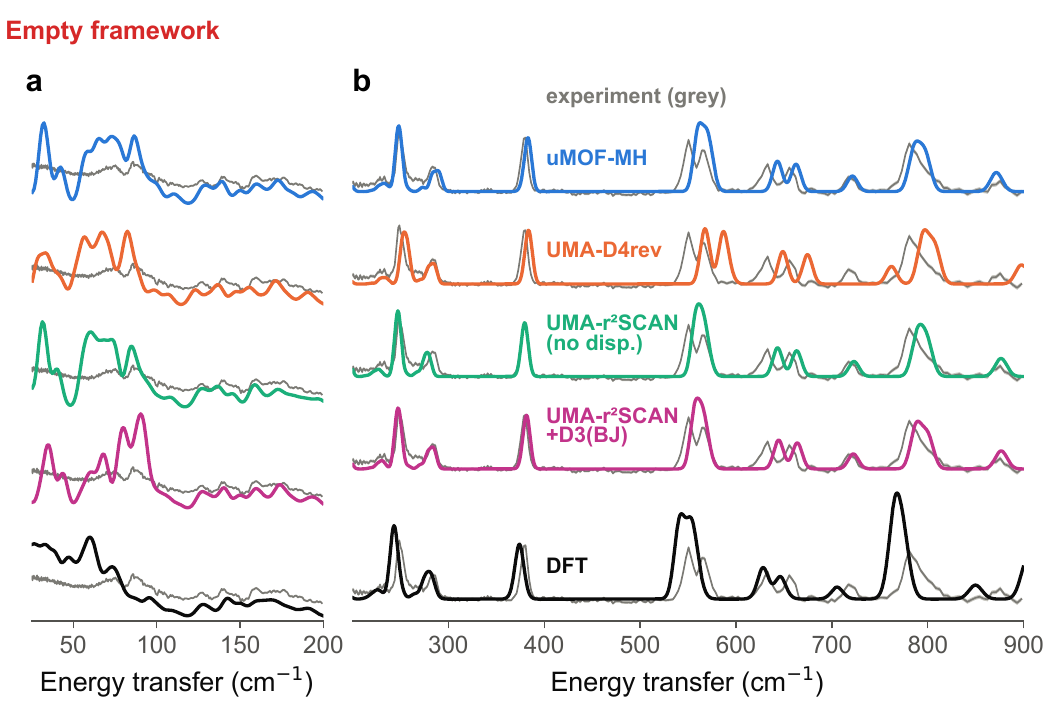}\\[2pt]
\includegraphics[width=\linewidth]{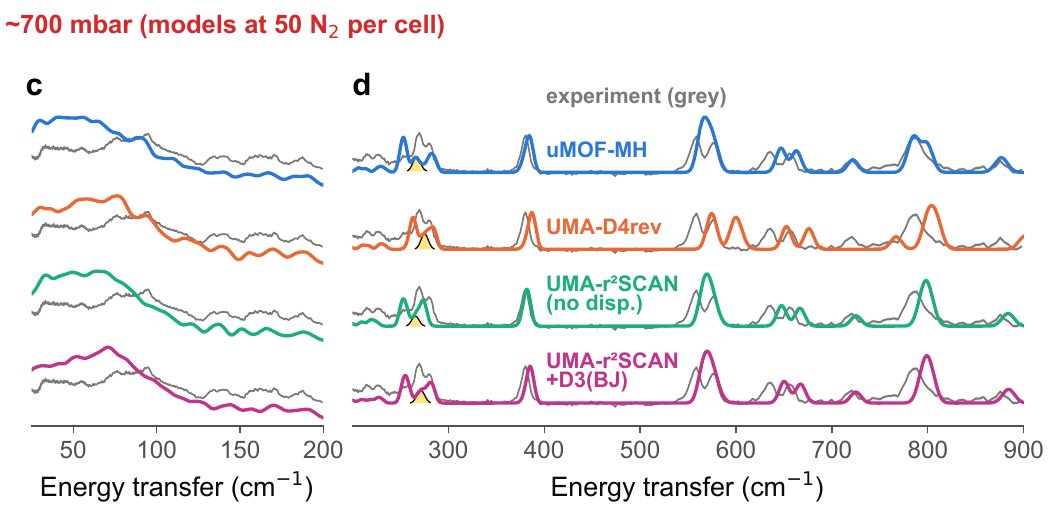}
\caption{Four MLIPs against experiment (grey), with no adjustment to the data or frequency scaling. (a, b) Empty framework, with DFT lattice dynamics (black) for comparison;\cite{Isotope2026} (c, d) the approximately 700~mbar state against each model at 50 \Ntwo\ per cell; yellow, the 270~\wn\ component of the three-band fit used for the experiment (Figure~\ref{fig:exp}c). (a, c) 25--200~\wn; (b, d) 200--900~\wn.}
\label{fig:models}
\end{figure}

We asked four frontier MLIPs to explain these changes, none fitted to these data: uMOF-MH, a MACE model fine-tuned to \rtwo-D4 data for gas-loaded MOFs,\cite{MACEMH,uMOF} and three UMA foundation models\cite{UMA} (UMA-D4rev, bare UMA-\rtwo\ and UMA-\rtwo+D3(BJ)\cite{D3,D3BJ}). With no frequency scaling or adjustment of any kind, all four reproduce the framework spectrum band for band from 200 to 900~\wn, empty and loaded (Figure~\ref{fig:models}). The three-band fit applied to the experiment (Figure~\ref{fig:exp}c) finds the new band in all four, at 265--275~\wn\ (Figure~\ref{fig:models}d, yellow): a separate peak in uMOF-MH and a shoulder in the others. For the empty framework they match the experiment as well as DFT lattice dynamics (PBE-D3(BJ); Figure~\ref{fig:models}a,b),\cite{Isotope2026} whose cost already limits what can be computed for this framework;\cite{Bahadur2025} for loaded cells the MLIPs stand alone. Agreement is weaker for the broad nitrogen signal at low energy, which comes from strongly anharmonic guest motion that the calculations may not represent well. What sets these calculations apart is speed. With uMOF-MH, a loaded cell is relaxed and its full force-constant matrix computed in about an hour on one GPU, no longer than it takes to measure a spectrum, and the spectra of both isotopologues then follow in minutes (SI Section~\ref{SI-si:calc}). Calculations of this kind ran during the beamtime, and at this speed a validated potential can explore, testing many configurations per loading, as we do below.

\begin{figure}[tbp]\centering
\includegraphics[width=0.9\linewidth]{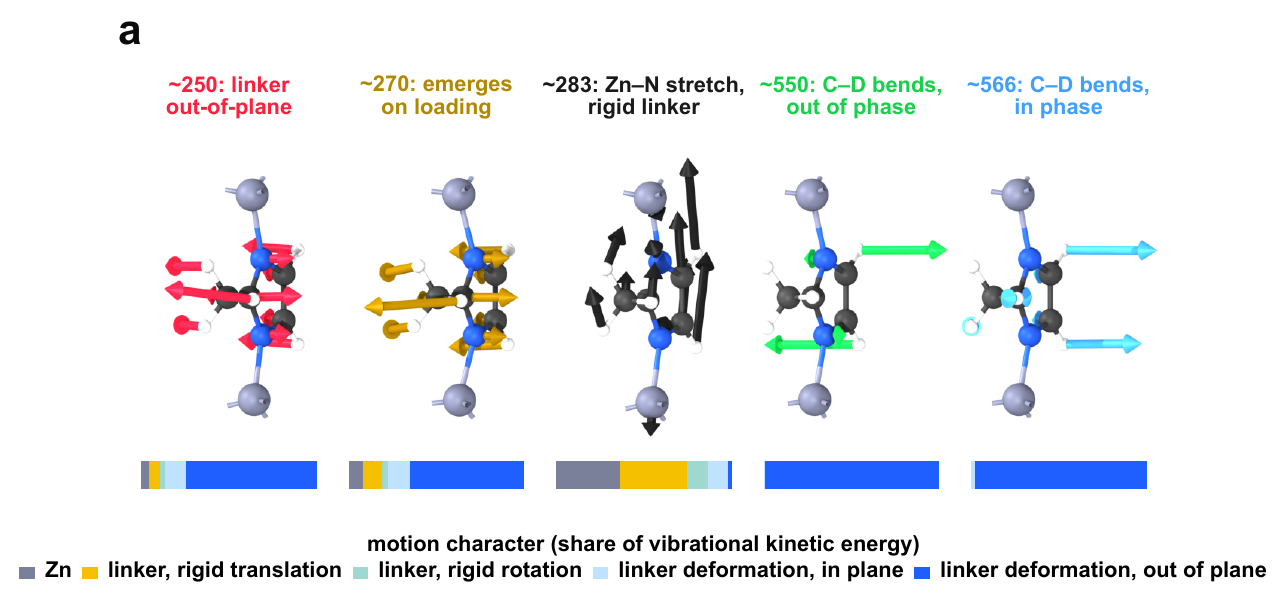}\\[-2pt]
\includegraphics[width=0.9\linewidth]{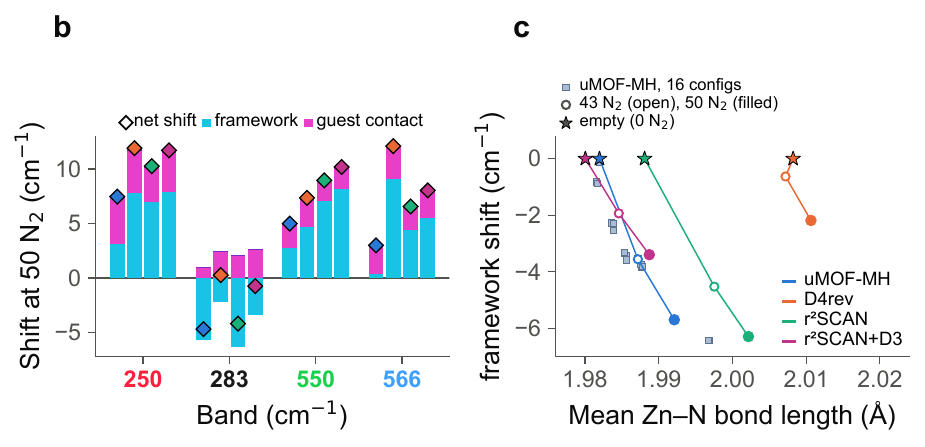}
\caption{Why the bands shift. (a) The tracked bands in the 50 \Ntwo\ cell (uMOF-MH; framework part of each mode, nitrogen not shown); bars, motion character of each band in the same cell. (b) Each shift split into dynamic (purple), guest-contact (magenta) and framework (cyan) terms, by changing one part of the force-constant matrix at a time (SI Figure~\ref{SI-fig:decomp}), for each model at 50 \Ntwo; diamonds, net shift, coloured by model as in c. (c) Framework term of the 283~\wn\ band against the mean Zn--N bond length.}
\label{fig:mech}
\end{figure}

Why do some bands stiffen and one soften? In the visualized eigenvectors (Figure~\ref{fig:mech}a), the stiffening bands are out-of-plane flexes of the linker faces that point into the pores, while the softening band is a Zn--N stretch. To explain these shifts in more detail, we split every shift exactly into three parts by changing one ingredient of the force-constant matrix at a time (SI Figure~\ref{SI-fig:decomp} and Section~\ref{SI-si:fcmath}). Holding the nitrogen still removes its motion but keeps its springs to the framework (the dynamic term); cutting those springs leaves the framework in its strained, loaded geometry (the guest-contact term); replacing that strained framework by the empty one gives the framework term. The answer is the same in all four potentials (Figure~\ref{fig:mech}b). Guest motion barely matters, and the nitrogen pressing on the pore-facing deuterons stiffens every band a little, but it is the framework itself that decides the direction of each shift. Adsorption squeezes the linker faces locally while stretching the network as a whole: where the cell expands on loading, the Zn--N bonds lengthen and the 283~\wn\ band softens, and the one potential whose cell contracts shifts it the wrong way (Figure~\ref{fig:mech}c; SI Section~\ref{SI-si:cellvolume}). The spectrum tells which potentials capture both effects.

\begin{figure}[tbp]\centering
\includegraphics[width=\linewidth]{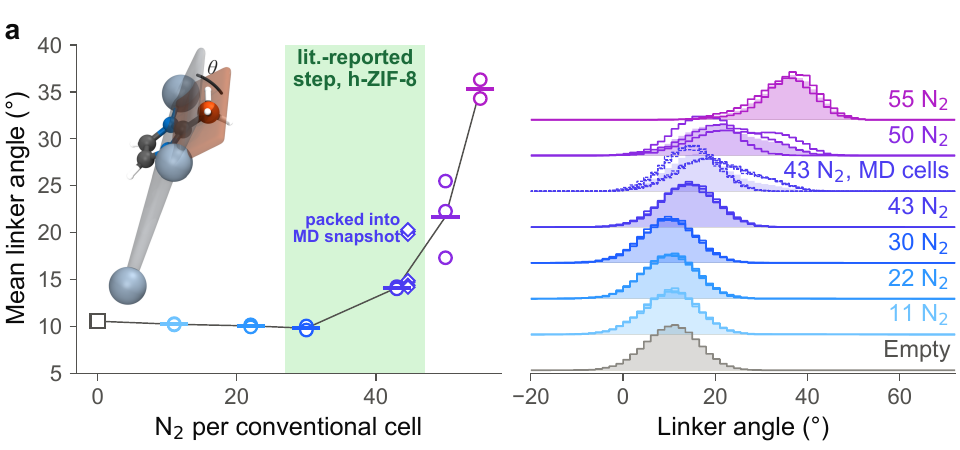}\\[2pt]
\includegraphics[width=0.5\linewidth]{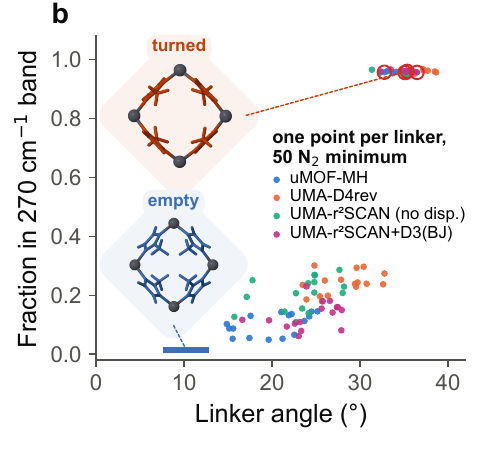}\includegraphics[width=0.5\linewidth]{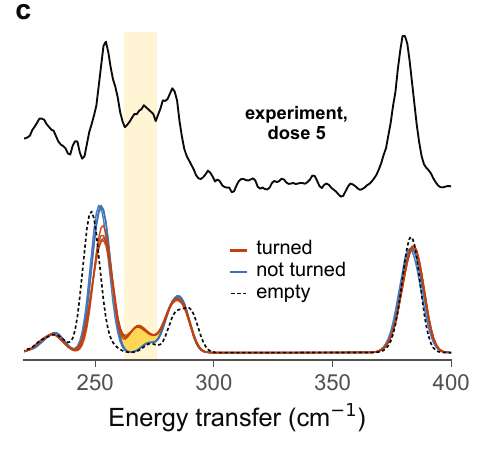}
\caption{The 270~\wn\ band counts turned linkers (uMOF-MH). (a) Mean linker angle against loading at 77~K (left; inset, the angle $\theta$) and its distributions (right); cells packed into the empty framework, except the 43 \Ntwo\ MD-framework cells. (b) Share of each linker's intensity in the 270 rather than the 250~\wn\ band against its angle, at 50 \Ntwo\ in each potential; insets, an empty and a turned four-ring (ringed points); bar, the empty-framework range. (c) Spectra of 43 \Ntwo\ cells with (red) and without (blue) a turned ring, the empty framework and experiment (dose~5); yellow, the new intensity near 270~\wn.}
\label{fig:linker}
\end{figure}

The emergence of a new band with dosing provides the most striking result. Molecular dynamics (MD) with uMOF-MH at 77~K shows that the linkers hardly respond to nitrogen up to 30 \Ntwo\ per cell, where the mean linker angle stays near 10$^\circ$, and then turn rapidly: to 14$^\circ$ at 43 \Ntwo, 22$^\circ$ at 50 and 35$^\circ$ at 55 (Figure~\ref{fig:linker}a). Above 30 \Ntwo\ the result also depends on where the molecules start: cells packed into an MD snapshot of the framework turn further than cells packed into the empty minimum (up to about 20$^\circ$ against 14$^\circ$ at 43 \Ntwo), and the three 50 \Ntwo\ runs span 17--26$^\circ$, so in a crowded framework the minimum reached depends on whether the linkers turn. The turning is collective: the four linkers around a four-membered Zn$_4$ window (a four-ring) turn together, to 29--38$^\circ$ against less than 23$^\circ$ for the rest, and no ring settles in between. Turning swings the linkers' deuterons out of the four-ring window and opens a hole through it (Figure~\ref{fig:linker}b, insets); the same swing opens the neighbouring six-ring gates. And it is these turned rings that produce the new band. On turned linkers, the 250~\wn\ out-of-plane flex moves up to about 270~\wn, in all four potentials, which put every turned linker at 31--39$^\circ$ (Figure~\ref{fig:linker}b). The intensity of the band is therefore a direct measure of how many rings have turned. Across 27 relaxed cells from 2 to 55 \Ntwo, the 270~\wn\ share rises in proportion to the fraction of turned linkers ($r=0.98$), while cells without turned linkers stay at the empty level whatever their loading (SI Figure~\ref{SI-fig:calib}). At a single loading the link is plain to see: of the 43 \Ntwo\ cells, those with a turned ring carry the 270~\wn\ band and those without do not (Figure~\ref{fig:linker}c). Applying this understanding of the peak to our experimental spectra shows that about an eighth of the linkers have turned by the third dose, and more than half by the sixth.

Deuteration is what makes this count possible. Calculated from the same force constants with hydrogen masses, the intensity from turned linkers merges with the Zn--N band near 295~\wn\ (SI Figure~\ref{SI-fig:HD}), and in a fully turned framework the two bands collapse into one, exactly as measured, repeatedly and by different groups, for hydrogenous ZIF-8 saturated with nitrogen (SI Figure~\ref{SI-fig:Hcheck}).\cite{Casco2016,Mor2025} In the deuterated framework, the same intensity stands alone in the valley, where it can be counted.

But why do the rings suddenly turn at these loadings? The answer is a single nitrogen molecule, threaded end-on through the window (SI Figure~\ref{SI-fig:threadevidence}a). In 41 trajectories below 55 \Ntwo, every turned ring held a molecule in its window at every check (395 of 395) and no unturned ring ever did (0 of 17\,989); in an unturned ring the window is closed off by the ring's own deuterons. Delete the molecule and the ring snaps back, in all three potentials tested, whereas deleting molecules from the cage wall changes nothing (SI Figure~\ref{SI-fig:threadevidence}b). The molecule props the window open (SI Figure~\ref{SI-fig:threadevidence}c and Section~\ref{SI-si:gate}). At 77~K a threaded window is held at 2.7--3.0~\AA\ in free diameter at its centre; in the 250 snapshots of empty windows we sampled, the widest opening was 2.2~\AA\ at 30 \Ntwo\ and 2.6~\AA\ at 50 \Ntwo, although rarer fluctuations will open it further, as we shall see below. The molecule holds the window open more permanently. Spectroscopically, the threaded molecule is everything. As noted above, of the ten cells at 43 \Ntwo\ only the four with a threaded, turned ring carry the 270~\wn\ band (Figure~\ref{fig:linker}c), yet the Zn--N band softens by much the same amount in all ten (SI Figure~\ref{SI-fig:threadevidence}e). One band reads how full the framework is; the other, how many windows have captured a molecule. This is the site located by in situ diffraction at 77~K\cite{Tian2016} and by crystallography under pressure,\cite{Hobday2018} close to the ``gate-blocked'' state inferred from methyl-rotor spectra.\cite{Titov2023}

\begin{figure}[tbp]\centering
\includegraphics[width=\linewidth]{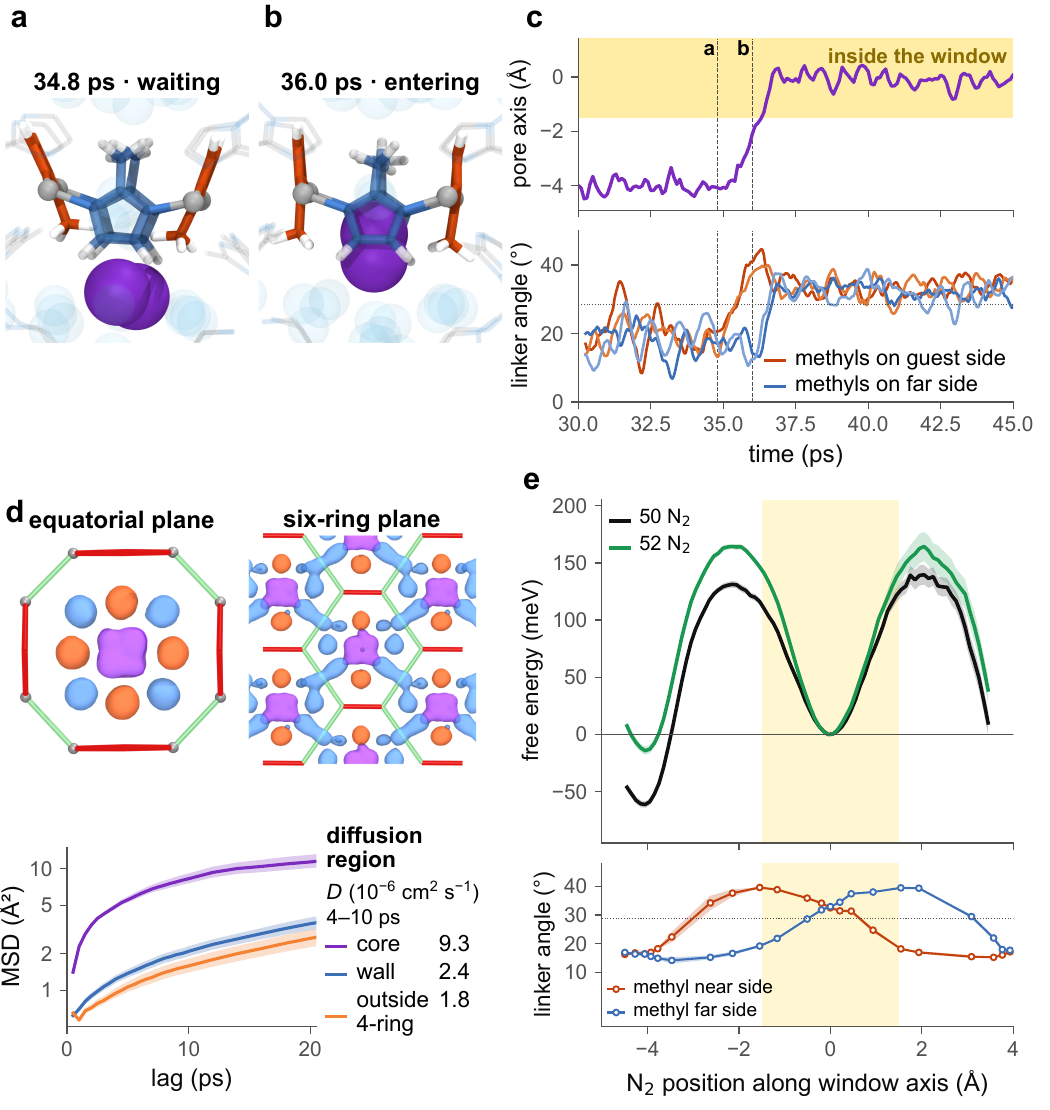}
\caption{A molecule threads a window on its own (uMOF-MH, 50 \Ntwo\ per cell, 200~K, unbiased MD; Movie~S1, ancillary file). (a, b) The four-ring and the threading \Ntwo\ (purple), waiting beside the closed window and entering once the two linkers whose methyl groups face it (orange) have swung open; blue, the far pair; window axis vertical. (c) Position of the molecule along the window axis (yellow, inside the window) and the four linker angles (0.5~ps running mean; dotted, turned threshold); dashed lines, the moments in a and b. (d) Density of the other molecules in two slices through a cage, by region: four-ring pocket (orange), core (purple) and wall (blue); four-ring windows red, six-ring green. Below, their mean-squared displacement by starting region, with apparent diffusion coefficients from 4--10~ps. (e) Free energy along the window axis from umbrella sampling at 200~K, relative to the threaded molecule, at 50 \Ntwo\ (black) and with two molecules added away from the four-rings (52 \Ntwo, green); below, the mean angle of the facing (orange) and far (blue) linker pairs at 50 \Ntwo. Bands: free energy, $\pm1$ standard deviation from a block bootstrap; linker angles, spread between the first and second halves of the sampling.}
\label{fig:threading}
\end{figure}

Threading the centre of a four-ring is quite rare on these timescales. A closed window excludes nitrogen entirely: no molecule ever sits between the ring and the shell of molecules waiting beside it (SI Section~\ref{SI-si:rare}). As the cages crowd, that shell fills and closes in on the window, moving onto its axis (SI Figures~\ref{SI-fig:approach} and \ref{SI-fig:dynamics}b), and the window breathes further open; but an empty window never opens as far as a threaded one. The waiting molecules are jammed in place, lying across the window rather than pointing into it (SI Figure~\ref{SI-fig:dynamics}), like guests pushing on the ``saloon doors'' seen by $^2$H NMR.\cite{Kolokolov2015} Threading therefore needs a coincidence: the linkers facing a waiting molecule must swing open just as it turns end-on.

At 77~K such a coincidence is too slow to catch in simulation, so we heated the cells to 200~K. In a single run, two molecules then threaded windows unaided: the first after 36~ps (Figure~\ref{fig:threading}a--c; SI Section~\ref{SI-si:event}), the second, at another window and from the opposite side, after 269~ps (SI Section~\ref{SI-si:event2}). Both followed the same path. Because the methyl groups alternate in direction around every four-ring, each side of a window has its own pair of methyl gatekeepers. In both events that pair swung wide open together in a large-amplitude fluctuation, the waiting molecule turned end-on and slid in within about 1~ps, and the far pair followed, leaving the ring turned. Neither molecule has left: the first stayed for the remaining 270~ps of the run, and for 50~ps after a copy of the cell was cooled to 77~K; the second, for the almost 40~ps since its entry. These times are set by how long we could afford to simulate, not by escape. The rest of the crowd shows why threading is so selective (Figure~\ref{fig:threading}d). The guests occupy pockets in front of the four-rings, a wall layer that includes the six-ring windows, and a core at the centre of each cage. The core molecules are the most mobile but stay in their cage; only wall molecules cross between cages, through the six-rings; and the pocket molecules, the least mobile of all, cross a four-ring only by threading it.

Umbrella sampling along the window (Figure~\ref{fig:threading}e) shows why the molecules stay: the threaded state is a free-energy minimum guarded by barriers of 0.13--0.16~eV on both sides, with the facing linker pair opening on the approach and the far pair as the molecule moves on (SI Section~\ref{SI-si:umbrella}). Whether a window prefers to be threaded is a closer call. At 200~K the threaded molecule lies 0.06~eV above the waiting one in free energy at 50 \Ntwo, and 0.01~eV above it when two molecules are added away from the four-rings (52 \Ntwo). Splitting the free energy into energy and entropy (SI Figure~\ref{SI-fig:energyentropy}) shows why: threading costs little or no energy at 50 \Ntwo\ and lowers it by about 0.07~eV at 52 \Ntwo, but pinning the molecule in the window costs it entropy, 0.07--0.09~eV at 200~K, about a third of it from locking the molecule end-on (SI Figure~\ref{SI-fig:orientation}). Crowding therefore favours threading, and cooling weakens the entropy penalty: extrapolated to 77~K, the balance tips towards the threaded state, consistent with cells at 55 \Ntwo\ with every ring turned lying 0.24--0.42~eV below those without (SI Section~\ref{SI-si:onset}). Such small differences are below the accuracy of current MLIPs, but the experiment settles the question: the 270~\wn\ band grows with loading, so in the real powder the balance does tip.

Each ring waits for its own threading event and, once threaded, stays turned, because at 77~K the surrounding crowd barely moves (SI Figure~\ref{SI-fig:dynamics}a) and cannot release it. This is why the band grows gradually rather than in a jump, and why the uptake turns non-exponential in the very doses in which the band grows fastest: the framework reorganizes slowly around its guests, one window at a time.

Fast MLIPs and a deuterated sample have turned a neutron spectrum into a molecule-by-molecule account of gas uptake in a flexible framework, with no fitting and with calculations that keep pace with the measurement. A band hidden in hydrogenous ZIF-8 counts the windows that single molecules have threaded and prised open: a site long known from diffraction is now seen, and counted, in a vibrational spectrum. The threaded state produces the band, is a free-energy trap guarded on both sides, and is reached by molecules on their own. The same combination, a deuterated host and potentials fast enough to explore while the beam is on, should expose hidden structural responses to guests in other flexible frameworks; and enhanced sampling, affordable at this speed, opens the way to a crowd that exchanges molecules with the gas.

\section*{Associated Content}
\textbf{Supporting Information.} Methods; experimental data, dosing balance and band tracking; calculations, low-energy descriptors and timings; derivation of the force-constant decomposition; band character and the Zn--N bond; linker angles, four-ring analysis, the 270~\wn\ calibration and hydrogenous spectra, including hydrogenous ZIF-8 saturated with nitrogen; construction of the loaded cells; window occupancy and apertures, the deletion and unloading tests and the onset of turning in three potentials; why threading is rare (approach to the window, window opening, mobility and orientation of the waiting molecules); the two threading events, the free energy, energy and entropy of threading, and nitrogen mobility.

\section*{Author Contributions}
J.A.: conceptualization, methodology, investigation, software, formal analysis, data curation, visualization, validation, project administration, writing -- original draft, writing -- review \& editing. H.C.: conceptualization (experiment proposal), writing -- review \& editing. P.S.: conceptualization (experiment proposal), resources (synthesis of the deuterated ZIF-8 sample), funding acquisition, writing -- review \& editing. M.E.P.: conceptualization (experiment proposal), funding acquisition, writing -- review \& editing.

\section*{Use of AI tools}
AI assistants (Claude, Anthropic; Codex, OpenAI) were used to help prepare figures, to write scripts for parts of the data analysis, and to draft large parts of the Supporting Information. The authors checked all results and text and take full responsibility for the content.

\section*{Acknowledgments}
We thank the ISIS Neutron and Muon Source for beamtime on TOSCA (experiment RB2610641).\cite{RB2610641} We are grateful for computational resources from STFC Scientific Computing's SCARF cluster.\cite{SCARF} This work has been supported by the EPSRC Fusion Grant 2022/27 [EP/W006839/1].

\section*{Data Availability}
The neutron scattering data from the dosing experiment are available through the ISIS data catalogue.\cite{RB2610641} Structures, force constants, simulated spectra and analysis scripts are available from the authors on reasonable request.

\clearpage
\setcounter{section}{0}\setcounter{figure}{0}\setcounter{table}{0}\setcounter{equation}{0}
\renewcommand{\thesection}{S\arabic{section}}\renewcommand{\thefigure}{S\arabic{figure}}\renewcommand{\thetable}{S\arabic{table}}\renewcommand{\theequation}{S\arabic{equation}}
\begin{center}{\Large\bfseries Supporting Information}\\[6pt]{\large Neutron Spectroscopy Transformed: Deuterated ZIF-8 and Machine-Learned Potentials Catch Nitrogen Threading Its Gates}\end{center}
\tableofcontents
\clearpage

\section{Experimental data}\label{SI-si:data}
Fully deuterated ZIF-8 (1.517~g) was dried under dynamic vacuum at about 120~$^\circ$C, dosed with nitrogen at 77~K from a 568.0~mL manifold, and measured on TOSCA\cite{Parker2014,Pinna2018,TOSCAinstrument} at ISIS at $15\pm2$~K after each dose. Data were reduced with Mantid 6.11\cite{Mantid} (graphite 002 analysers, monitor normalization, the same mask of 36 detector tubes for every state) on TOSCA's native energy bins. The same measurement of the empty sample can (TSC33878--33886) was subtracted as background from every spectrum, so the background errors of different states are correlated. The framework itself was measured without nitrogen twice: a long measurement (TSC33670--33682) used for the four-model comparison, and a shorter one at the start of the dosing series used as the reference for the doses (series runs TSC36917--37040; Table~\ref{SI-tab:series}).

\begin{table}[htbp]\centering\small
\begin{tabular}{ll}\toprule
State & Runs (TSC)\\\midrule
No nitrogen & 36917--36924 (8)\\
Dose 1 & 36927--36937 (11)\\
Dose 2 & 36939--36947 (9)\\
Dose 3 & 36949--36991 (43)\\
Dose 4 & 36993--37001 (9)\\
Dose 5 & 37003--37014 (12)\\
Dose 6 & 37016--37024 (9)\\
Approx.\ 700~mbar & 37032--37040 (9)\\\bottomrule
\end{tabular}
\caption{Runs combined for each state of the dose series (equal-weight means; number of runs in brackets).}\label{SI-tab:series}
\end{table}

Comparisons between states normalize locally (band tracking, Section~\ref{SI-si:tracking}) or scale each loaded spectrum to the one without nitrogen over 200--900~\wn.

\subsection{Dosing balance}\label{SI-si:dosing}
The apparent cumulative loading after dose $j$, in \Ntwo\ per conventional cell, is
\begin{equation}
L_j^{\rm app}=\frac{V_{\rm w}}{R\,n_{\rm cell}}\sum_{i=1}^{j}\frac{P_{i,\rm fill}-P_{i,\rm close}}{T_i},
\end{equation}
where $V_{\rm w}=0.568$~L is the volume of the manifold at room temperature, $n_{\rm cell}=0.532$~mmol the amount of conventional cells in the sample, $R$ the gas constant, $P_{i,\rm fill}$ the manifold pressure before dose $i$ was opened to the sample, $P_{i,\rm close}$ the pressure when it was closed, and $T_i$ the manifold temperature during dose $i$. This gives 10.7, 21.5, 30.4, 36.6, 43.0 and 49.5 \Ntwo\ per cell. The pressure fall measures gas leaving the manifold, which includes gas in the cell free volume, on external surfaces and in mesopores, so these are upper bounds on pore occupancy. The closing pressures (3.9--6.5~mbar after 30~min for doses 1--4, 7.8 and 27.6~mbar for doses 5 and 6) are operational endpoints, not equilibrium pressures. Gauge, volume and temperature uncertainties have not been established. The approximately 700~mbar state was reached by admitting nitrogen until the pressure was approximately steady.

\subsection{Band tracking}\label{SI-si:tracking}
Six band families were defined on the empty spectrum (238--300~\wn\ with three components; 365--398, 536--590, 617--672, 706--738 and 760--812~\wn) and fitted with non-negative Gaussians of shared width on a linear baseline, by weighted least squares on the unsmoothed data. States were fitted in dose order, each starting from the previous fit, with components allowed to move by at most 7~\wn\ per step. Errors are from the covariance scaled by $\sqrt{\chi^2_\nu}$ and omit reduction and background uncertainty. The central share is $A_2/(A_1+A_2+A_3)$ of the 238--300~\wn\ family; its bootstrap uncertainty (400 refits) is 0.013--0.023. Moving the window edges by $\pm3$~\wn\ and the starting positions by $\pm2$~\wn\ changes it by at most 0.019. The share rises above the empty level (0.068) from dose~2 (0.143), and every step after dose 3 is resolved. A third component in this family is favoured over two from dose~2 on ($\Delta$BIC $-22$ to $-290$), and shifts of $\pm1$ and $\pm3$~\wn\ injected into the dose-3 spectrum are recovered to within 0.5~\wn.

\section{How the simulations were performed}\label{SI-si:howsims}
All calculations used the conventional cell of ZIF-8 (276 framework atoms) and ran in double precision on single A100 GPUs of the SCARF cluster;\cite{SCARF} MD used ASE.\cite{ASE} 

\subsection{The potentials}\label{SI-si:potentials}
Four MLIPs were used: the MACE model uMOF-MH and three UMA models (UMA-D4rev, bare UMA-\rtwo\ and UMA-\rtwo+D3(BJ)). Unless stated otherwise, the loaded cells, the molecular dynamics and the free-energy sampling use uMOF-MH, and the UMA models are compared with it where stated. uMOF-MH was not trained or fitted on the present spectra; its training data include gas-loaded MOFs.\cite{uMOF} The UMA models are \texttt{uma-s-1p2p1-ft-d4rev} (\texttt{omol} task) and \texttt{uma-s-1p2p1-ft-r2scan} (\texttt{omat} task).

uMOF-MH was fine-tuned from MACE-MH-1 on \rtwo-D4 energies, forces and stresses,\cite{uMOF} so its training targets include D4 dispersion. The bare UMA-\rtwo\ and UMA-\rtwo+D3(BJ) pair differ only in the added dispersion (D3(BJ), $s_9=0$), but each structure was relaxed with its own model, so the comparison includes the effect of dispersion on the minimum. UMA-D4rev is a different checkpoint, not a D4-on/off control.

\subsection{Timings}\label{SI-si:calc}
Timings come from 15 uMOF-MH configurations (11--55 \Ntwo) that each went through 30~ps of MD at 77~K, relaxation and $6n_a$ central-displacement force evaluations on one A100 GPU (Table~\ref{SI-tab:times}). Every one of these minima has residual forces below $10^{-4}$~eV\,\AA$^{-1}$ and no imaginary modes; its lowest mode apart from the three translations lies at 5.0--14.5~\wn. The summed active time is 3.3--5.2~h per configuration, and 3.7--6.4~h from the start of MD to finished spectra including scheduling gaps; relaxation and force constants alone take 0.8--2.4~h (median 1.3~h).

\begin{table}[htbp]\centering\small
\begin{tabular}{rcrrrrr}\toprule
\Ntwo\ per cell & Cell & MD (h) & Relaxation + phonons (h) & Spectra (s) & Total (h) & Wall clock (h)\\\midrule
11 & 2 & 1.901 & 1.956 & 57 & 3.873 & 4.553 \\
22 & 2 & 2.087 & 1.299 & 54 & 3.401 & 3.720 \\
30 & 1$^{*}$ & 3.168 & 1.461 & 57 & 4.645 & 6.352 \\
30 & 2 & 2.229 & 2.444 & 58 & 4.690 & 5.404 \\
30 & 3 & 2.224 & 1.346 & 97 & 3.598 & 4.354 \\
37 & 1$^{*}$ & 3.244 & 1.441 & 60 & 4.702 & 6.353 \\
37 & 3 & 2.353 & 1.349 & 59 & 3.718 & 3.903 \\
43 & 1 & 3.763 & 1.044 & 62 & 4.823 & 6.353 \\
43 & 2 & 2.517 & 1.301 & 109 & 3.848 & 5.864 \\
43 & 3 & 2.480 & 0.820 & 101 & 3.328 & 4.138 \\
50 & 1 & 3.893 & 1.051 & 64 & 4.961 & 6.354 \\
50 & 2 & 2.706 & 2.107 & 65 & 4.831 & 4.880 \\
50 & 3 & 2.630 & 0.831 & 104 & 3.490 & 4.163 \\
55 & 1 & 4.006 & 1.166 & 66 & 5.190 & 6.354 \\
55 & 2 & 2.927 & 1.080 & 113 & 4.038 & 5.744 \\
\bottomrule\end{tabular}
\caption{Time taken for each loaded cell (uMOF-MH, one A100 GPU). MD, 30~ps at 77~K; spectra, both isotopologues on a CPU. Total, the sum of the three steps; wall clock, from the start of MD to the finished spectra, including waiting in the queue. $^{*}$Cell preparation restored from an earlier run, whose time is not included.}\label{SI-tab:times}
\end{table}

\subsection{Building the simulations}\label{SI-si:operations}
Every simulation is a sequence of the operations of Table~\ref{SI-tab:operations}. Loadings are \Ntwo\ per conventional cell.

{\small\setlength{\tabcolsep}{4pt}\renewcommand{\arraystretch}{1.2}
\begin{longtable}{@{}>{\raggedright\arraybackslash}p{0.2\linewidth}>{\raggedright\arraybackslash}p{0.76\linewidth}@{}}
\caption{The operations from which every simulation is built.}\label{SI-tab:operations}\\
\toprule Operation & Meaning\\\midrule\endfirsthead
\toprule Operation & Meaning\\\midrule\endhead
\bottomrule\endfoot
Packing & Molecules inserted one at a time, each at the lowest-energy of 16 random trial positions, framework and cell fixed; after each insertion only the guests are relaxed. (The older loading-scan runs and the 49 \Ntwo\ cell relaxed all atoms after each insertion.)\\\midrule
Addition & One molecule inserted into a running MD state at the emptiest point, nothing relaxed (gradual dosing only).\\\midrule
Deletion & Named molecules removed.\\\midrule
Full relaxation & Atoms, then atoms and all cell parameters at zero pressure, to $10^{-4}$~eV\,\AA$^{-1}$. Used for every minimum whose spectrum is compared with experiment.\\\midrule
NVT MD & Fixed cell, Nos\'e--Hoover chain, 77~K unless stated; 0.25~fs step for 2~ps, then 0.5~fs.\\\midrule
Phonons and spectrum & $\Gamma$-point force constants from $\pm0.005$~\AA\ central displacements; AbINS TOSCA spectrum at 10~K, deuterium masses.\cite{AbINS,Euphonic} Spectra are never computed from MD frames.\\
\end{longtable}}

\subsection{The two families of loaded cells}\label{SI-si:families}
Every loaded cell was built in the same way: nitrogen was packed into a fixed framework, the cell was run for 30~ps of NVT MD at 77~K with new velocities, and it was then relaxed. The cells fall into two families that differ only in the framework the nitrogen was packed into. \emph{Empty-framework cells} were packed into the relaxed empty framework. \emph{MD-framework cells} were packed into snapshots of the framework taken from MD of cells already holding 38 or 39 \Ntwo, with that nitrogen removed, so that their linkers start from positions adapted to a crowded pore. At 43 \Ntwo, half of the MD-framework cells have a turned four-ring and none of the empty-framework cells do. For the comparison of the four potentials, one cell was taken from each family, a 43 \Ntwo\ MD-framework cell with one turned four-ring and a 50 \Ntwo\ empty-framework cell with two, and every potential was relaxed from exactly these structures.

\subsection{Spectra and normalization}\label{SI-si:norm}
Spectra are $\Gamma$-point AbINS spectra for TOSCA at 10~K with fundamentals only;\cite{AbINS} for the deuterated framework the 120 framework hydrogen masses are replaced by deuterium masses with the force constants unchanged. Below 200~\wn\ every curve is normalized to unit area over 25--200~\wn. Above 200~\wn\ each model is scaled once so that its band near 380~\wn\ matches the experimental height above a local baseline; no frequency scaling or shift is applied. In main-text Figures~2 and 4 the model spectra above 200~\wn\ are rebuilt from the $\Gamma$-mode sticks at a resolution matched to the measured band widths (Gaussian FWHM 3.5~\wn\ plus 1.25\% of the energy transfer); at the AbINS width the sticks reproduce the AbINS curves ($r=0.993$--0.998), and the 270~\wn\ fit is unchanged (265.7--275.4~\wn).

\section{Band character and the shift decomposition}\label{SI-si:fcmath}\label{SI-si:modes}

\subsection{Motion character}
For the INS-weighted band modes of the empty cell, the kinetic energy of each linker is split into a rigid-body part, fitted per linker (translation, and rotation about the N$\cdots$N axis, the ring normal and the in-plane perpendicular), and internal deformation, divided into in-plane and out-of-plane parts relative to the ring plane; the rest is carried by Zn. The split is the same in all four potentials to within 0.03. Main-text Figure~3a shows the same split for the loaded bands of the 50 \Ntwo\ cell (modes within $\pm6$~\wn\ of each pictured mode, INS-weighted). It barely differs from the empty-cell bands, and the new 270~\wn\ band has the character of the 250~\wn\ out-of-plane flex (65\% out-of-plane deformation) with more Zn and rigid-linker motion; the nitrogen carries at most 0.3\% of the kinetic energy.

\subsection{Following a band}
Bands were followed from the empty to the loaded cell, and from the deuterated to the protonated framework, by collective eigenvector following.\cite{Isotope2026} A band is the set of empty-cell modes within $\pm6$~\wn\ of the model's own fitted band position, weighted by the INS weight $w_j=\sum_a\sigma_a|e_{aj}|^2/m_a$. Its loaded position is the overlap-weighted centroid of the loaded modes within $\pm40$~\wn, with overlaps $O_{jk}=(e_j\cdot e_k)^2$ computed on the framework atoms after rigid translation is removed and methyl deuterons relabelled (methyl groups rotate between minima). Varying the band half-width (4--8~\wn) and the search window (30--50~\wn) changes the shifts by at most 1.5~\wn, and fitting each model's simulated spectrum instead gives shifts of the same sign in all 24 cases (Table~\ref{SI-tab:si-shifts}).

\begin{table}[htbp]\centering\footnotesize\setlength{\tabcolsep}{4pt}
\begin{tabular}{llrrrrrr}\toprule
Band & Model & 43 \Ntwo & 50 \Ntwo & Framework & Guest contact & Dynamic & Zn--N part\\
 & & \multicolumn{2}{c}{total shift} & \multicolumn{4}{c}{split of the 50 \Ntwo\ shift}\\\midrule
250~\wn\ (expt +6.1) & uMOF-MH & +6.1 & +7.5 & +3.1 & +4.2 & +0.1 & +0.0\\
 & UMA-D4rev & +9.1 & +11.9 & +7.8 & +3.8 & +0.3 & +0.8\\
 & bare UMA-\rtwo & +8.2 & +10.3 & +7.0 & +3.1 & +0.2 & +0.3\\
 & UMA-\rtwo+D3 & +9.2 & +11.7 & +7.8 & +3.7 & +0.2 & +0.4\\
\midrule
283~\wn\ (expt $-3.3$) & uMOF-MH & $-2.9$ & $-4.7$ & $-5.7$ & +1.0 & +0.0 & $-4.5$\\
 & UMA-D4rev & +1.0 & +0.3 & $-2.2$ & +2.4 & +0.1 & $-2.4$\\
 & bare UMA-\rtwo & $-2.9$ & $-4.2$ & $-6.3$ & +2.1 & +0.0 & $-5.6$\\
 & UMA-\rtwo+D3 & +0.3 & $-0.7$ & $-3.4$ & +2.6 & +0.1 & $-3.3$\\
\midrule
550~\wn\ (expt +7.6) & uMOF-MH & +2.8 & +5.0 & +2.8 & +2.2 & +0.0 & +0.1\\
 & UMA-D4rev & +4.9 & +7.3 & +4.7 & +2.6 & +0.0 & +0.0\\
 & bare UMA-\rtwo & +5.9 & +9.0 & +7.1 & +1.8 & +0.0 & +0.1\\
 & UMA-\rtwo+D3 & +6.7 & +10.2 & +8.2 & +2.0 & +0.0 & +0.1\\
\midrule
566~\wn\ (expt +10.3) & uMOF-MH & +1.3 & +3.0 & +0.3 & +2.6 & +0.0 & +0.0\\
 & UMA-D4rev & +8.5 & +12.1 & +9.1 & +3.0 & +0.0 & +0.0\\
 & bare UMA-\rtwo & +3.6 & +6.6 & +4.4 & +2.1 & +0.0 & +0.1\\
 & UMA-\rtwo+D3 & +4.7 & +8.0 & +5.5 & +2.5 & +0.0 & +0.0\\
\bottomrule
\end{tabular}
\caption{Eigenvector-followed band shifts (\wn) from each model's empty cell, and the exact split of the 50 \Ntwo\ shift (Section~\ref{SI-si:modes}). Framework and guest-contact terms are the mean of the two evaluation orders; their division is uncertain by at most 0.3~\wn\ (0.6~\wn\ for the 250~\wn\ band). The Zn--N part is the portion of the framework term reproduced by the 48 Zn--N spring blocks alone. Experimental shifts: empty to approximately 700~mbar.}\label{SI-tab:si-shifts}
\end{table}

\subsection{The decomposition}
The loaded force-constant matrix $\Phi$ is split into framework (f) and guest (g) blocks. The guest springs on framework atom $a$ are $\mathbf{K}_a=\mathrm{sym}(-\sum_g\Phi_{ag})$, the part of the framework--guest coupling that the acoustic sum rule places on the framework diagonal. The band is followed through four matrices (Figure~\ref{SI-fig:decomp}): (a) the full loaded $\Phi$; (b) the framework block $\Phi_{ff}$ alone, which is exactly the limit of infinitely heavy, fixed nitrogen; (c) $\Phi_{ff}$ with the guest springs subtracted from its diagonal, i.e.\ the strained loaded framework on its own; and (d) the empty framework, its atoms matched one to one. The dynamic, guest-contact and framework terms are the differences (a)$-$(b), (b)$-$(c) and (c)$-$(d), and they sum exactly to the total shift. The three rigid translations are projected out before diagonalization.

The dynamic term is small because the framework bands (250--570~\wn) lie far above the nitrogen frequencies (below 120~\wn): it is at most 0.3~\wn\ in all four potentials. The guest-contact term is positive for every band (1--4~\wn), as springs that resist motion stiffen a mode. To first order it is $\sum_a \mathbf{e}_a^{\mathsf T}\mathbf{K}_a\mathbf{e}_a/m_a$, so it is large for modes that move the pore-facing ring deuterons along the directions in which the nitrogen presses on them: the deuterons supply 63--80\% of it in the out-of-plane linker bands (250, 550, 566~\wn). The Zn--N stretch barely moves those atoms, so its guest-contact term is only $+1.0$~\wn\ (uMOF-MH), and the softening of the framework's own Zn--N springs ($-5.7$~\wn) decides the sign of its shift. The framework term includes any change in framework--framework coupling caused by the nitrogen through the many-body potential as well as the change in geometry.

Strain and guest springs could also be applied in the other order, adding the springs (rotated into each linker's empty-cell orientation) to the empty framework before straining it. We report the mean of the two orders; they differ by at most 0.3~\wn\ (0.6~\wn\ for the 250~\wn\ band) at 50 \Ntwo.

\sifigure{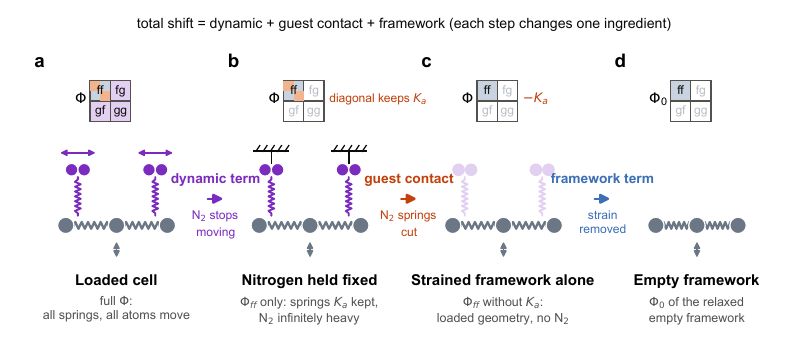}{The four force-constant matrices of the decomposition: (a) the loaded cell; (b) nitrogen held fixed, which removes the dynamic term; (c) guest springs removed, which removes the guest-contact term and leaves the strained framework; (d) the empty framework. Differences between successive matrices give the terms of main-text Figure~3b.}{SI-fig:decomp}

\subsection{The Zn--N bond}
The Zn--N stretching force constant $k=-\hat b\cdot\Phi_{\mathrm{Zn,N}}\cdot\hat b$, averaged over the 48 Zn--N bonds of the cell, falls on loading to 50 \Ntwo\ by 5.6\% (uMOF-MH), 2.4\% (UMA-D4rev), 6.7\% (bare UMA-\rtwo) and 3.4\% (UMA-\rtwo+D3(BJ)), as the mean bond lengthens by 10.1, 2.4, 14.1 and 8.7~m\AA; across all models and cells it falls by 0.4--0.55\% per m\AA. Exchanging only the 48 Zn--N spring blocks between the empty and loaded frameworks reproduces 78--100\% of the 283~\wn\ framework term at 50 \Ntwo\ while changing the other bands by at most 0.1~\wn\ (Figure~\ref{SI-fig:znn}). At 33--39 \Ntwo\ (uMOF-MH) the 283~\wn\ framework term ($-0.8$ to $-2.5$~\wn) is not yet carried by the Zn--N part ($+0.3$ to $-0.7$~\wn), so other framework changes dominate at moderate loading.

\sifigure[.6]{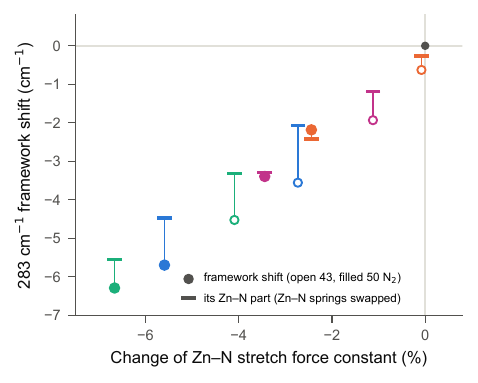}{Framework term of the shift of the 283~\wn\ band (circles; open 43 \Ntwo, filled 50 \Ntwo) and its Zn--N part (bars) against the change in the Zn--N stretching force constant, for the four potentials.}{SI-fig:znn}

\subsection{Cell volume against loading}\label{SI-si:cellvolume}
The comparison of the four potentials uses one relaxed structure per loading, and the loaded UMA-D4rev cells are smaller than its empty cell. To test whether this belongs to the potential or to how the cells were built, nitrogen was added to UMA-D4rev's own empty minimum one molecule at a time, each at the lowest-energy of 16 trial positions, with a full relaxation of atoms and cell at zero pressure after every addition, in three independent series (Figure~\ref{SI-fig:cellvolume}). The cell contracts from the first molecules on, by 1--3~\AA$^3$ per molecule (from 4919.5 to 4871--4897~\AA$^3$ at 15--25 \Ntwo), and the mean Zn--N bond shortens steadily; no ring has turned. Relaxing the earlier 0, 43 and 50 \Ntwo\ structures from the uMOF-MH volume or from one scaled to the UMA-D4rev empty volume reaches the same minima (within 1.3~\AA$^3$ and 0.02~eV). The contraction is therefore a property of UMA-D4rev, consistent with its 283~\wn\ band shifting the opposite way to the other potentials. uMOF-MH cells relaxed after 77~K MD stay within about 1\% of the empty volume up to 43 \Ntwo\ and expand above it.

\sifigure{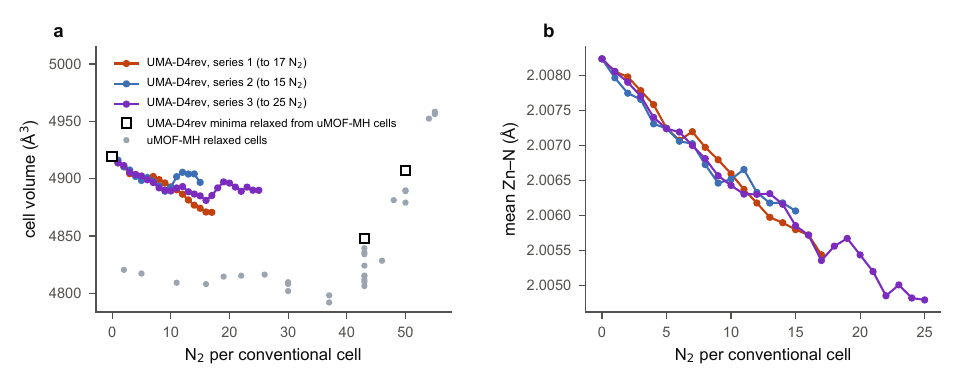}{Cell volume (a) and mean Zn--N bond length (b) against loading. Lines: UMA-D4rev, one molecule added at a time with a full zero-pressure relaxation after each addition (three independent series); squares, the earlier UMA-D4rev minima relaxed from uMOF-MH structures; grey, uMOF-MH cells relaxed after 30~ps of MD at 77~K.}{SI-fig:cellvolume}

\section{Linker rotation and the 270~\wn\ band}\label{SI-si:linker}

\subsection{Trajectories and coordinate}
The loading series of main-text Figure~4a uses the empty-framework cells (Section~\ref{SI-si:families}): three at each of 11, 22, 30, 43 and 50 \Ntwo\ and two at 55, run at 77~K in the fixed cell of the empty minimum and analysed over $5<t\le30$~ps. The empty reference is a separate 77~K run of the empty framework (125--375~ps). The eight MD-framework cells at 43 \Ntwo\ use the same protocol; four of them each have one turned four-ring.

The linker angle is a Zn--Zn--Zn--methyl-C dihedral about the Zn--Zn axis that the linker bridges (main-text Figure~4a, inset). Each linker lies in two six-rings; the context giving the smaller angle in the DFT reference structure was chosen once, with its sign, and angles are folded into $[-90^\circ,90^\circ)$.

\subsection{Turning happens ring by ring}\label{SI-si:fourrings}
Each linker belongs to exactly one of the six four-rings of the cell, so the 24 dihedrals split into six groups of four. Across all 42 trajectories (252 rings; Figure~\ref{SI-fig:fourrings}) the ring means are bimodal: no ring lies between 23.1 and 28.6$^\circ$ except transiently, while a molecule threads it. Unturned rings follow the slow uniform rotation of the cell (about 10$^\circ$ up to 34 \Ntwo, 14$^\circ$ at 43, 17--19$^\circ$ at 49--50); turned rings lie at 29--38$^\circ$. Cells hold no turned ring up to 41 \Ntwo, 0 or 1 at 43 (four of the eight MD-framework cells and none of the three empty-framework cells), 0--2 at 50 and all six at 55. At 77~K turned rings are turned within the first 0.5~ps and never change identity, whereas the uniform rotation develops over a few picoseconds and is reversible.

\sifigure{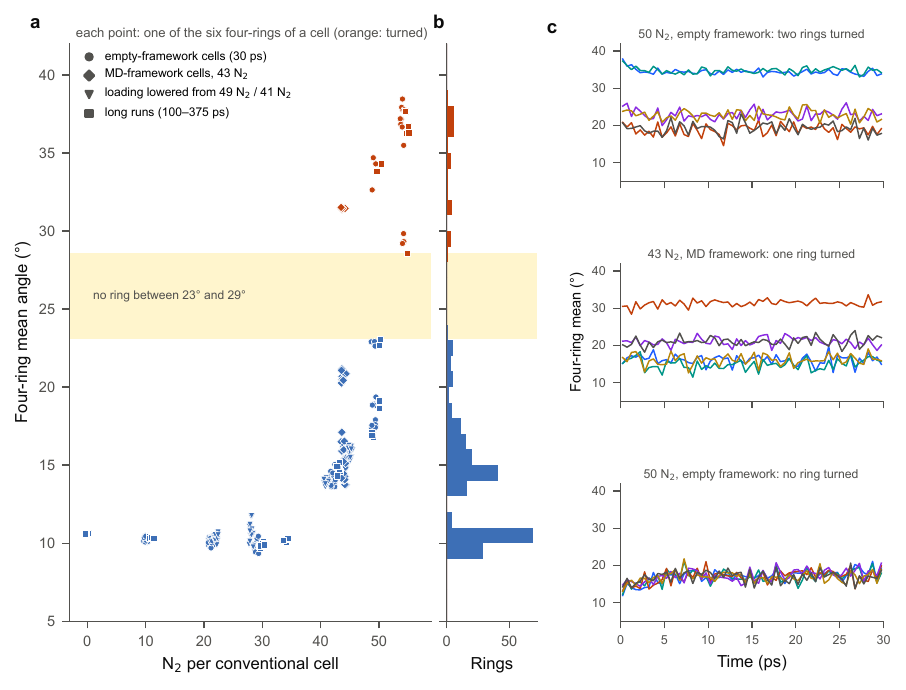}{Linker rotation resolved into four-rings (uMOF-MH, 77~K). (a) Mean dihedral of each four-ring for every trajectory (orange, turned; blue, unturned; yellow band, the empty interval between 23.1 and 28.6$^\circ$). (b) Histogram of all 252 ring means. (c) Ring means against time for a 50 \Ntwo\ cell with two turned rings, an MD-framework cell at 43 \Ntwo\ with one, and a 50 \Ntwo\ cell with none.}{SI-fig:fourrings}

\subsection{The 270~\wn\ band counts turned linkers}\label{SI-si:calib}
Per-linker shares of a band are the INS-weighted kinetic-energy shares of its loaded modes. In the 50 \Ntwo\ minimum of each potential (main-text Figure~4b), each linker's share of the 270~\wn\ band rises with its angle (rank correlation 0.78--0.82), and the six most-rotated linkers carry 55--68\% of the band. The loaded 270~\wn\ modes draw 36--61\% of their character from the empty 250~\wn\ band and 2--35\% from the 283~\wn\ band; nitrogen carries only 0.1--0.3\% of their kinetic energy.

For the calibration (Figure~\ref{SI-fig:calib}) the central share from the same three-component fit used for experiment is compared with the fraction $f$ of turned linkers (angle at least 28.6$^\circ$) in 27 relaxed uMOF-MH cells from 2 to 55 \Ntwo. Twenty cells, up to 50 \Ntwo, have no turned linker and stay at the empty level. The fit is $\mathrm{share}=0.091+0.408f$ ($r=0.98$, residual SD 0.028). Inverted for the measured shares it gives turned fractions of about 0.04 (dose~1), 0.13 (doses 2 and 3), 0.31, 0.38 and 0.60 (doses 4--6), and 1.0 at 700~mbar, with prediction ranges of about $\pm0.1$. These depend on one potential's intensities and are illustrative.

\sifigure[.55]{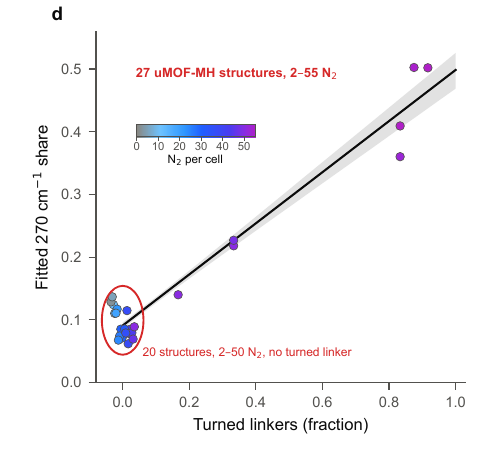}{The 270~\wn\ share against the fraction of turned linkers for 27 relaxed uMOF-MH cells from 2 to 55 \Ntwo, coloured by loading; line, fit ($r=0.98$).}{SI-fig:calib}

\subsection{Deuteration separates the turned-linker band}\label{SI-si:HD}
The same force constants give the spectrum of the hydrogenous framework when the framework deuterium masses are replaced by hydrogen (Figure~\ref{SI-fig:HD}). With deuterium, turned rings produce a separate band at 266--269~\wn, in the valley between the 252 and 284~\wn\ bands, growing with the number of turned rings. With hydrogen all bands move up (the 250~\wn\ flex to about 280~\wn, the Zn--N stretch to about 295~\wn), and the turned-ring intensity falls on the upper side of the existing 295~\wn\ band: no new band appears.

Measured INS of hydrogenous ZIF-8 dosed with nitrogen at 77~K to 0.5 and 0.8~bar (data of ref~\citenum{Mor2025}) shows exactly this (Figure~\ref{SI-fig:Hcheck}): the 278~\wn\ band disappears and the intensity collapses into one band at 299.5~\wn. Calculated with hydrogen masses, cells without turned rings keep two bands, cells with one or two turned rings begin to merge them, and cells with all six turned give a single band at 293.5~\wn; with deuterium masses the same cells keep two bands. The saturated hydrogenous framework thus behaves as a fully turned one, an independent check on an isotopologue not used anywhere else in this work.

\sifigure{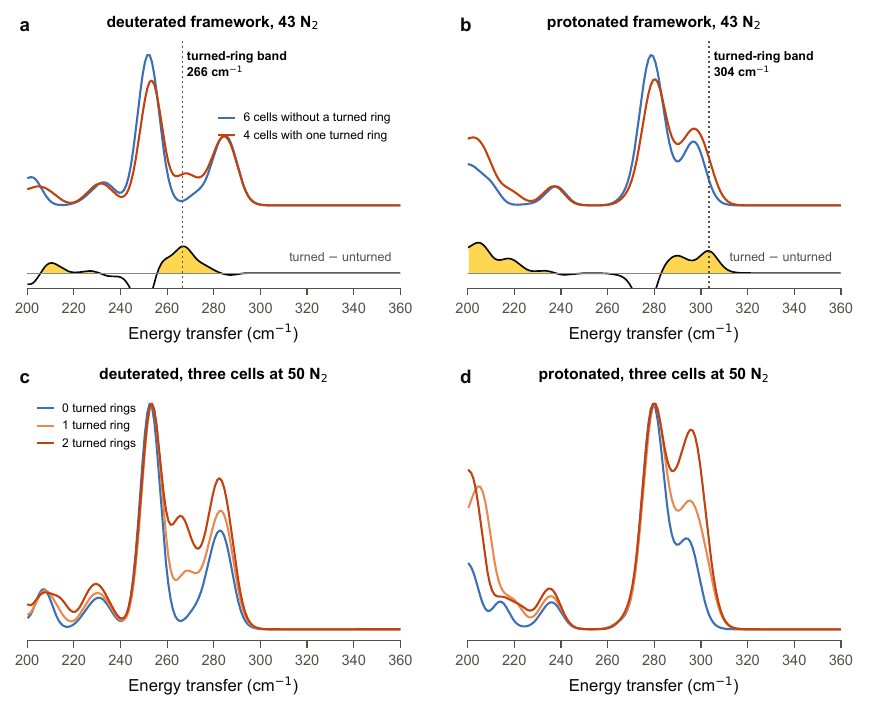}{Framework partial spectra (AbINS, TOSCA, 10~K) of uMOF-MH cells with deuterium (a, c) or hydrogen (b, d) masses. (a, b) Ten cells at 43 \Ntwo: mean of the four with a turned ring (orange) and of the six without (blue); below, their difference (yellow). (c, d) Three cells at 50 \Ntwo\ with zero, one and two turned rings.}{SI-fig:HD}

\sifigure{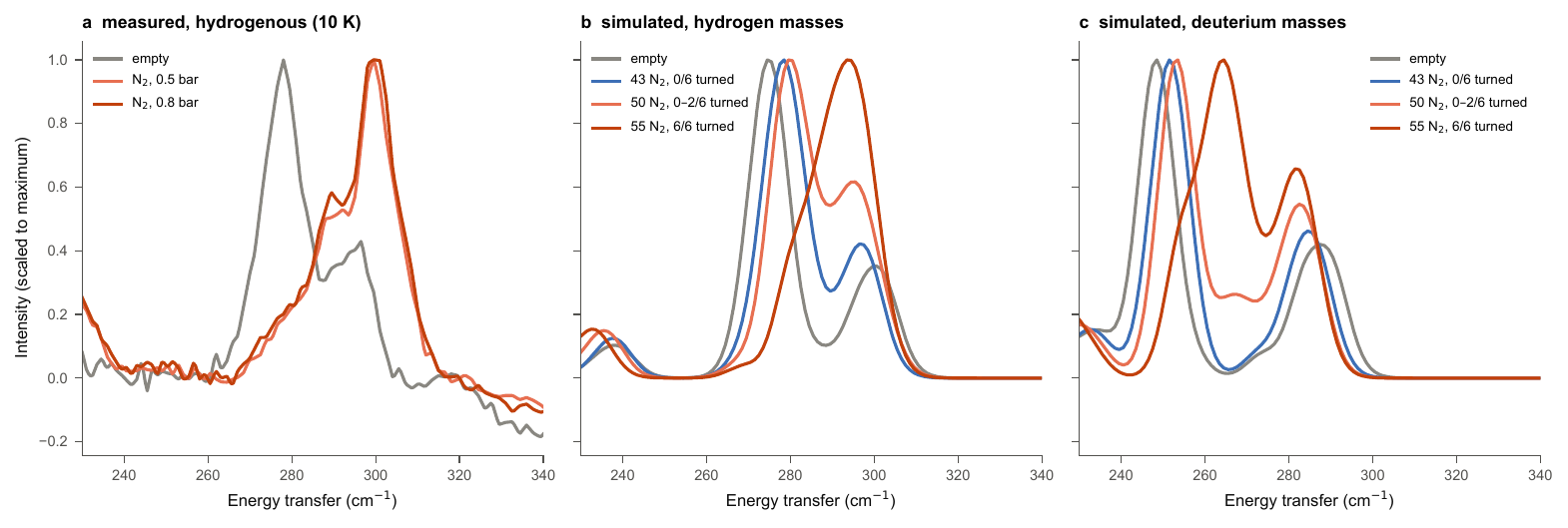}{Hydrogenous ZIF-8 with nitrogen. (a) Measured INS at 10~K, empty and after dosing to 0.5 and 0.8~bar at 77~K (data of ref~\citenum{Mor2025}; linear baseline removed, scaled to maximum). (b) uMOF-MH spectra with hydrogen masses: empty, 43 \Ntwo\ (no turned ring), 50 \Ntwo\ (up to two turned) and 55 \Ntwo\ (all six turned). (c) The same cells with deuterium masses.}{SI-fig:Hcheck}

\section{The window molecule turns the ring}\label{SI-si:trap}
Figure~\ref{SI-fig:threadevidence} collects the evidence, summarized in the main text, that a molecule threaded through a four-ring window holds the ring turned; the subsections below give the details.

\sifigure{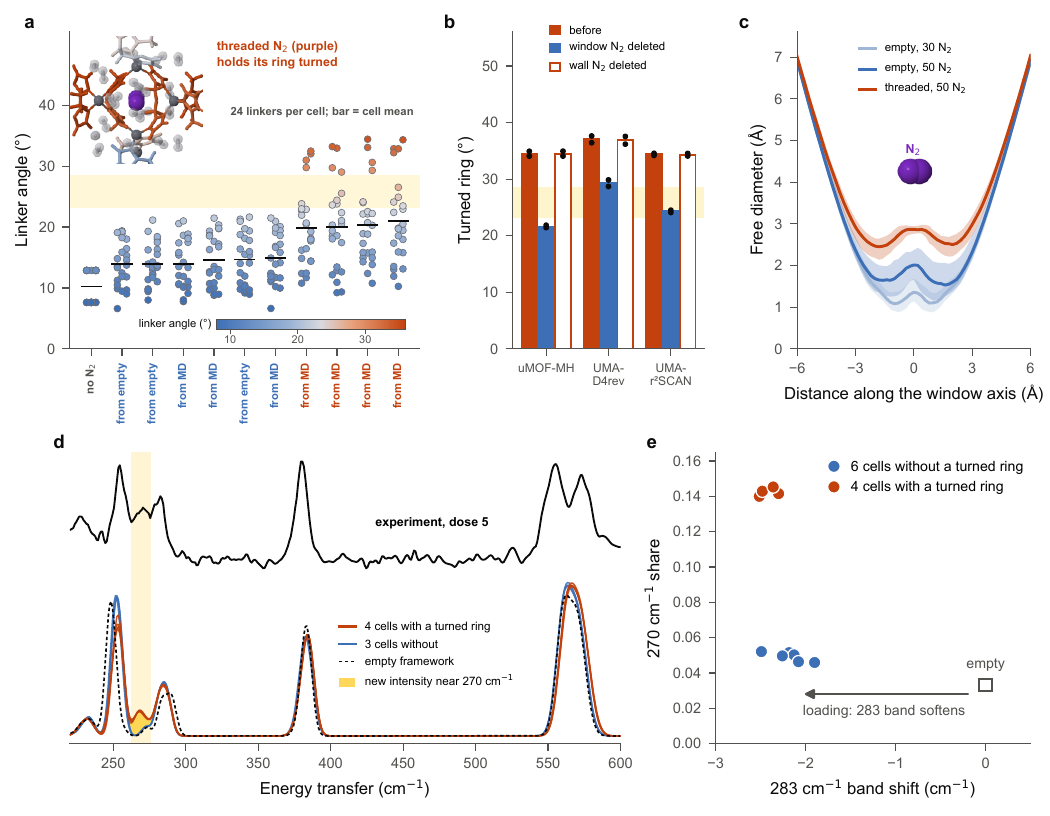}{Evidence that a threaded molecule turns a four-ring (uMOF-MH unless stated). (a) Linker angles of ten relaxed cells at 43 \Ntwo, packed into the empty framework or an MD snapshot, with (orange labels) and without (blue) a turned ring; inset, a threaded \Ntwo\ (purple) in a turned ring at 50 \Ntwo; yellow band, the range no ring mean settles in. (b) Turned rings before and after deleting the window or wall molecules, three potentials. (c) Free diameter along the four-ring axis, 77~K MD: median and 10--90\% range for empty windows at 30 and 50 \Ntwo\ and threaded windows at 50 \Ntwo. (d) Spectra of the seven MD-framework cells, the empty framework and experiment (dose~5), as main-text Figure~4c. (e) For each cell, the 270~\wn\ share against the shift of the 283~\wn\ band.}{SI-fig:threadevidence}

\subsection{Where the nitrogen sits}\label{SI-si:cagesites}
Molecular centres from the 77~K trajectories were mapped onto the two cages of the cell. Up to 43 \Ntwo\ almost all molecules sit in a layer 3--6~\AA\ from the cage centre (19.9 per cage at 43 \Ntwo; Figure~\ref{SI-fig:cagesites}a), with about one per cage at the centre from 37 \Ntwo. As the loading rises the wall layer locks into a lattice of fixed sites (Figure~\ref{SI-fig:cagesites}c,d): a molecule visits four to six sites in 25~ps at 11 \Ntwo\ but only one from about 43 \Ntwo.

Four-ring windows are occupied only at turned rings. In 41 trajectories with zero, one or two turned rings, every turned ring held a molecule in its window in every 1~ps block (395 of 395) and no unturned ring ever did (0 of 17\,989; Figure~\ref{SI-fig:cagesites}b). The window molecule lies along the ring normal ($|\cos\theta|=0.96$--0.98). In unturned rings the ring deuterons lie 1.6--1.9~\AA\ from the window centre and close it. The result holds for window radii of 1.0--2.5~\AA.

\sifigure[.8]{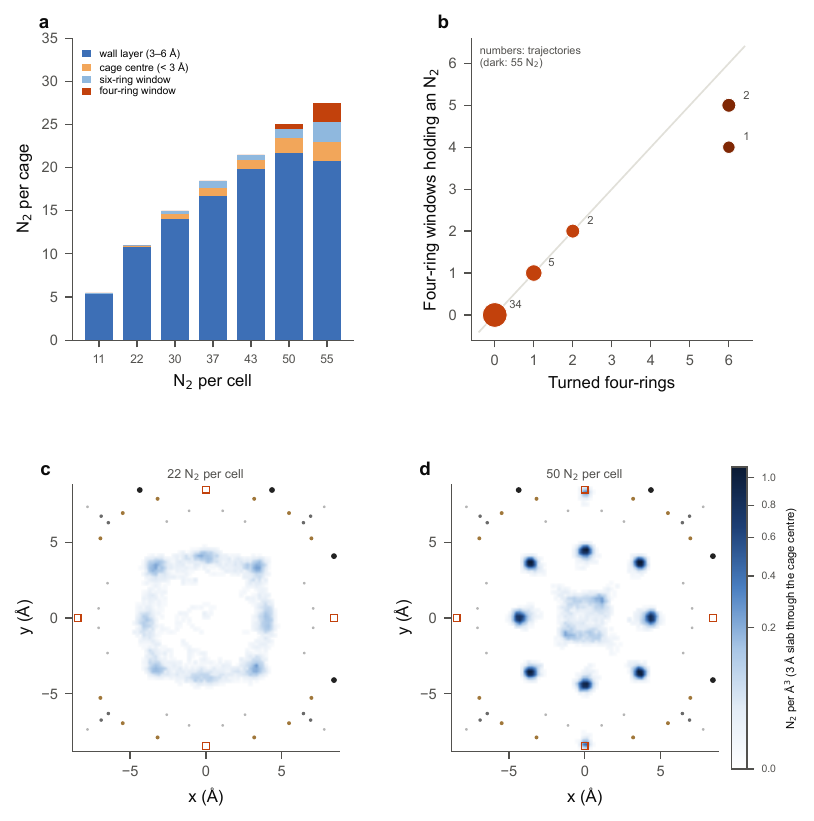}{Where the nitrogen sits (uMOF-MH, 77~K MD). (a) Molecules per cage in each region. (b) Four-ring windows holding a molecule against turned four-rings, per trajectory (numbers: trajectories at each point). (c, d) Density of molecular centres in a 3~\AA\ slab through a cage centre at 22 and 50 \Ntwo, with framework atoms (dots) and four-ring windows (squares).}{SI-fig:cagesites}

\subsection{Deleting the window nitrogen}\label{SI-si:windowtests}
Deleting the window molecules from relaxed minima and relaxing again turns the rings back (Figure~\ref{SI-fig:windowtests}a). In the uMOF-MH 50 \Ntwo\ minimum the two turned rings fall from 34.1 and 34.9$^\circ$ to 21.4 and 21.9$^\circ$; deleting only one molecule turns back only its ring; in two MD-framework cells at 43 \Ntwo\ the turned ring falls from 33.1 to 20.9$^\circ$ and from 31.2 to 18.8$^\circ$. Deleting two wall-layer molecules instead changes nothing (34.1 and 35.0$^\circ$). UMA-D4rev and bare UMA-\rtwo\ behave the same way (window deletion: 36.5/37.6 to 28.7/29.9$^\circ$ and 34.2/34.6 to 24.5/24.1$^\circ$; wall deletion: below 0.5$^\circ$). In 30~ps of 77~K MD after the deletion every ring stays unturned (Figure~\ref{SI-fig:windowtests}b), and the spectrum loses the 270~\wn\ band: its share falls from 0.207 to 0.073, against 0.030 for the empty framework (Figure~\ref{SI-fig:windowtests}c).

\sifigure{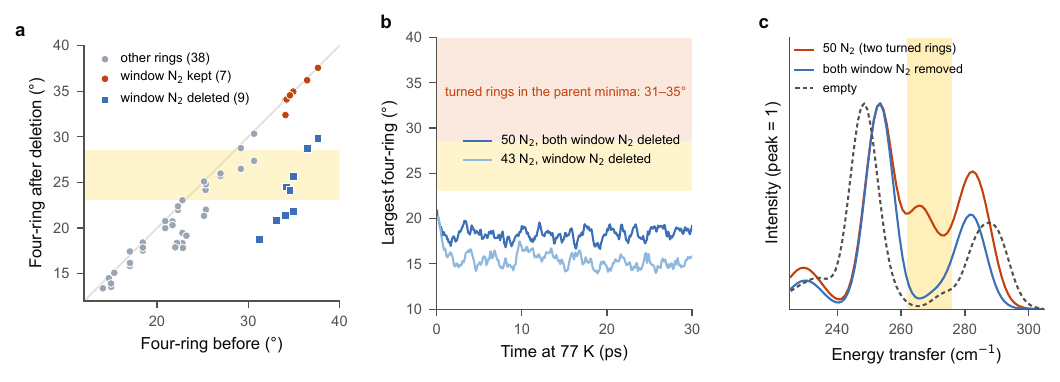}{Deleting the window nitrogen. (a) Every four-ring before and after deletion from relaxed minima (uMOF-MH; UMA-D4rev and bare UMA-\rtwo): rings whose window molecule was deleted (blue squares), rings whose molecule was kept (orange) and all other rings (grey); yellow band, 23.1--28.6$^\circ$. (b) 77~K MD after deletion: largest of the six four-ring means. (c) Spectra (deuterated framework) of the 50 \Ntwo\ minimum, the same cell with both window molecules removed and the empty framework; yellow, the window of the 270~\wn\ share.}{SI-fig:windowtests}

\subsection{Unloading and gradual dosing}\label{SI-si:dosingsim}
Removing wall and centre molecules from the 50 \Ntwo\ minimum down to 30 \Ntwo\ while keeping both window molecules leaves the two rings turned (33--34$^\circ$ over 30~ps at 77~K); removing the window molecules as well turns them back (14--15$^\circ$). Conversely, in three chains that received one molecule every 10~ps at 77~K from 37 to 55 \Ntwo\ no ring turned (largest ring mean 22--24$^\circ$ at 55 \Ntwo), whereas cells packed directly at 55 \Ntwo\ have all six rings turned. Turning follows the window molecule, not the loading as such.

\subsection{Onset of turning in three potentials}\label{SI-si:onset}
MD-framework cells rebuilt with UMA-D4rev and bare UMA-\rtwo\ (three each at 37 and 43 \Ntwo) turn no ring at 37 \Ntwo; at 43 \Ntwo\ two of three UMA-D4rev cells and all three bare UMA-\rtwo\ cells turn one ring, each with its window occupied, against four of eight with uMOF-MH (Figure~\ref{SI-fig:onset}a). The onset is similar in the three potentials. In further uMOF-MH cells, most cells at 46 \Ntwo\ and every cell at 48 and 52 \Ntwo\ keep at least one turned ring. At 55 \Ntwo\ the packed cells with all six rings turned lie 0.24--0.42~eV below the gradual-dosing minima in the same cell (Figure~\ref{SI-fig:onset}c). In all these cells the rings that turn are those whose window held a molecule when packing finished (43 of 44, against 5 of 130 with an empty window, all of which ended inside the gap): packing can place a molecule in a closed window because the framework is held fixed during insertion, and on relaxation the ring opens around it.

\sifigure{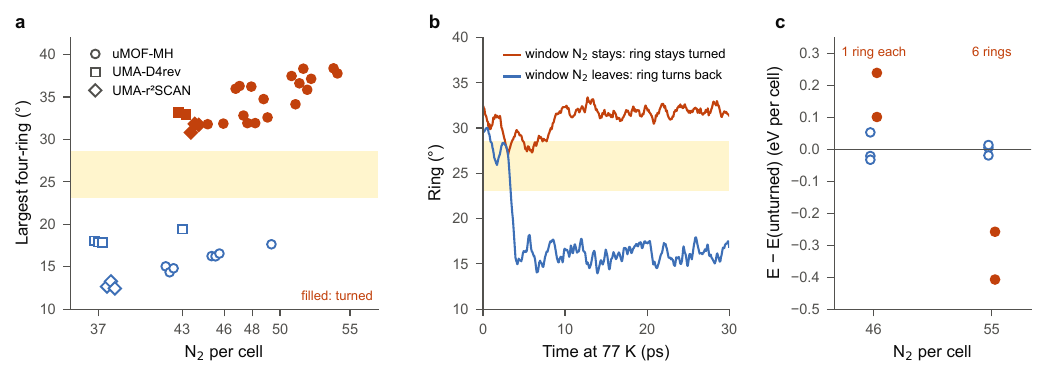}{Onset of ring turning. (a) Largest four-ring mean (last 20~ps of 30~ps at 77~K) of every cell against loading: uMOF-MH (circles), UMA-D4rev (squares) and bare UMA-\rtwo\ (diamonds); filled, turned; yellow band, 23.1--28.6$^\circ$. (b) Two 46 \Ntwo\ cells in which a ring starts turned: its window molecule stays (orange) or leaves (blue). (c) Relaxed energies relative to the unturned cells at the same loading: 46 \Ntwo\ (one turned ring) and 55 \Ntwo\ (six turned rings, against cells loaded one molecule at a time).}{SI-fig:onset}

\section{Threading: why it is rare and how it happens}\label{SI-si:rare}
The first three subsections look, in the 77~K MD, at what stands between a molecule beside a closed four-ring window and the threaded state; the rest describe the two threading events at 200~K and the free energy of threading.

\subsection{The approach to a window}\label{SI-si:approach}
No nitrogen centre ever lies 1--3.5~\AA\ from a window centre: that space is the ring itself (Figure~\ref{SI-fig:approach}). A threaded molecule sits within 0.75~\AA\ of the centre; the next molecules form a shell 3.5--5.25~\AA\ away, one on each cage side, which fills from 1.0 molecules per window at 22 \Ntwo\ to about 2.0 at 50--55 \Ntwo\ and moves closer (mean 4.64 to 4.15~\AA).

\sifigure{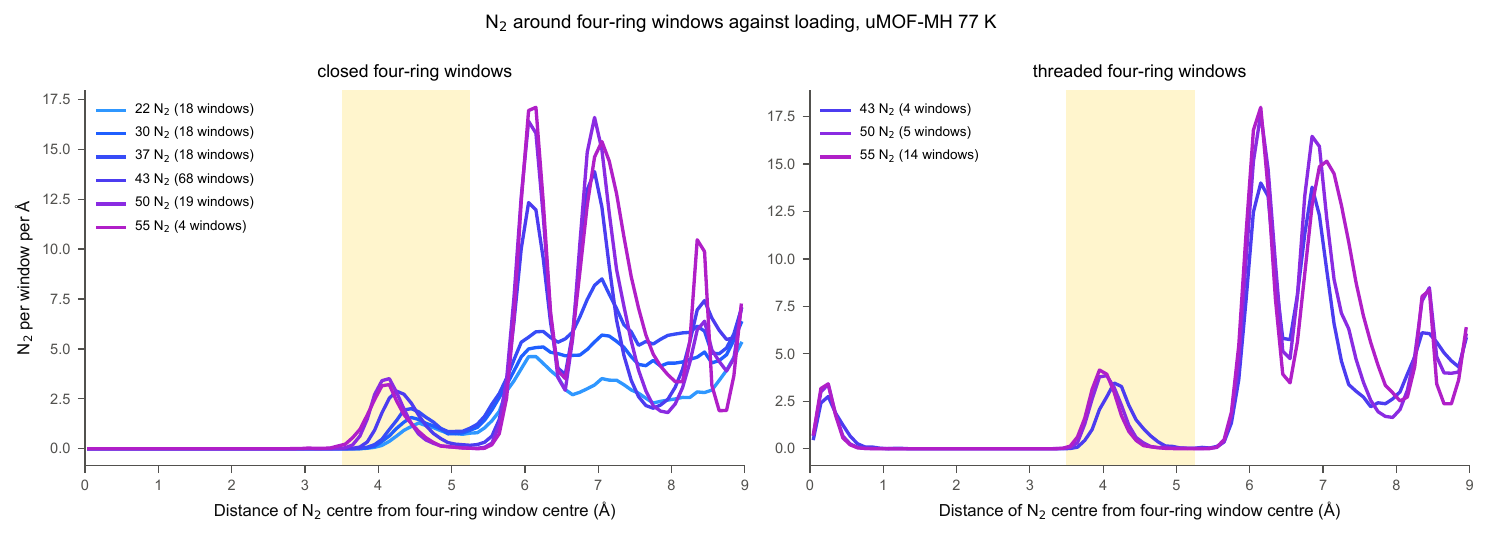}{Nitrogen centres against distance from four-ring window centres (uMOF-MH, 77~K): unoccupied windows at 22--55 \Ntwo\ (left) and threaded windows at 43--55 \Ntwo\ (right). Yellow, the shell beside the window.}{SI-fig:approach}

\subsection{How far does the window open?}\label{SI-si:gate}
At each height $s$ along the window axis ($s=0$ in the plane of the four Zn atoms), Figure~\ref{SI-fig:gate} gives the radius of the largest sphere that can be centred within 1.6~\AA\ of the axis without overlapping any framework atom. Other molecules never enter $|s|<2$~\AA, so the profile is set by the framework alone. The narrowest points lie about 1~\AA\ either side of the ring plane, at the ring deuterons. The narrowest free radius of unoccupied windows grows with loading, from a median of 0.49~\AA\ at 30 \Ntwo\ to 0.73~\AA\ at 50 \Ntwo; turned rings without a molecule reach 0.91~\AA, and threaded windows are held at 1.30~\AA. At the window centre the free diameter of unoccupied windows reaches at most 2.2~\AA\ at 30 \Ntwo\ and 2.6~\AA\ at 50 \Ntwo\ (150 and 100 snapshots), against 2.7--3.0~\AA\ for threaded windows. The window thus breathes further open as the cages fill, but a ring that has turned without a molecule is still about 0.8~\AA\ narrower in diameter than a threaded one: the molecule completes the opening.

\sifigure{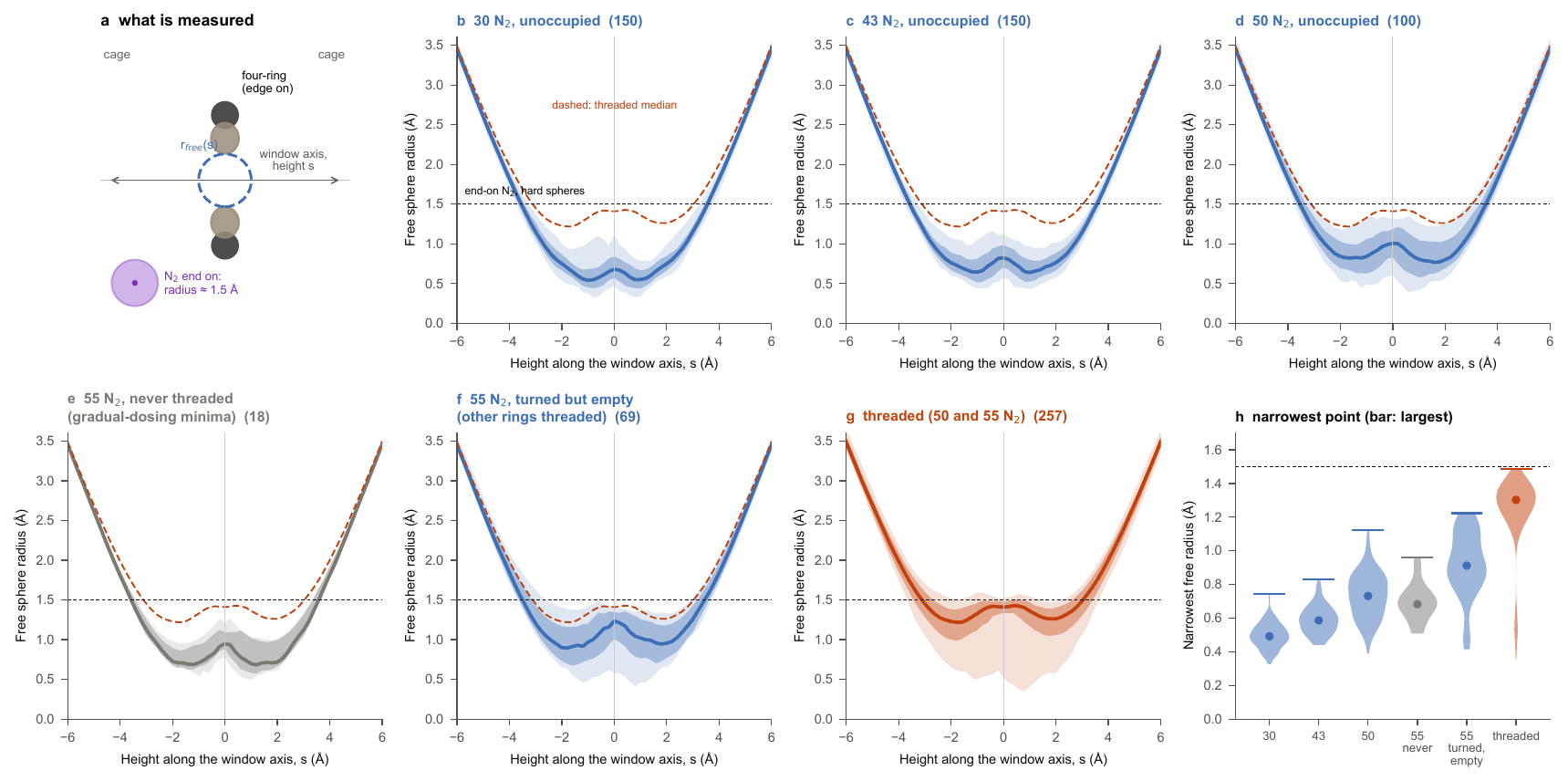}{How far a four-ring window opens. (a) The quantity measured: the largest free sphere at height $s$ along the window axis. (b--g) Framework-only profiles for unoccupied windows at 30, 43 and 50 \Ntwo, never-threaded windows at 55 \Ntwo\ (gradual-dosing minima), turned but empty windows at 55 \Ntwo, and threaded windows at 50 and 55 \Ntwo; line, median; bands, 10--90\% and full range; dotted, end-on \Ntwo. (h) Narrowest free radius of every window snapshot.}{SI-fig:gate}

\subsection{The waiting molecules}\label{SI-si:waiting}
From about 41 \Ntwo\ the molecules barely move at 77~K: in the runs of 100--375~ps at 43--55 \Ntwo\ their mean-squared displacement stays below 1~\AA$^2$ for as far as the statistics are sound (tens to about 100~ps), against 17--38~\AA$^2$ at 11--34 \Ntwo, and cage changes are rare (11 in about 131 molecule-nanoseconds; Figure~\ref{SI-fig:dynamics}a). Those beside a closed four-ring do not creep at all (0.35~\AA$^2$), and threaded molecules are the least mobile (0.20--0.23~\AA$^2$). The waiting molecules move onto the window axis as the loading rises (95\% within 1~\AA\ of it at 43 \Ntwo, 100\% at 50--55 \Ntwo), but they lie across the window: only 1.2--2.3\% point within 20$^\circ$ of the window centre, against 6\% for random orientation (Figure~\ref{SI-fig:dynamics}b). Threading therefore needs a coincidence: the ring must breathe open just as the waiting molecule turns end-on and pushes in. At 77~K this is too rare to see in our simulations; at 200~K it happened twice in one run (Sections~\ref{SI-si:event} and \ref{SI-si:event2}).

\sifigure[.8]{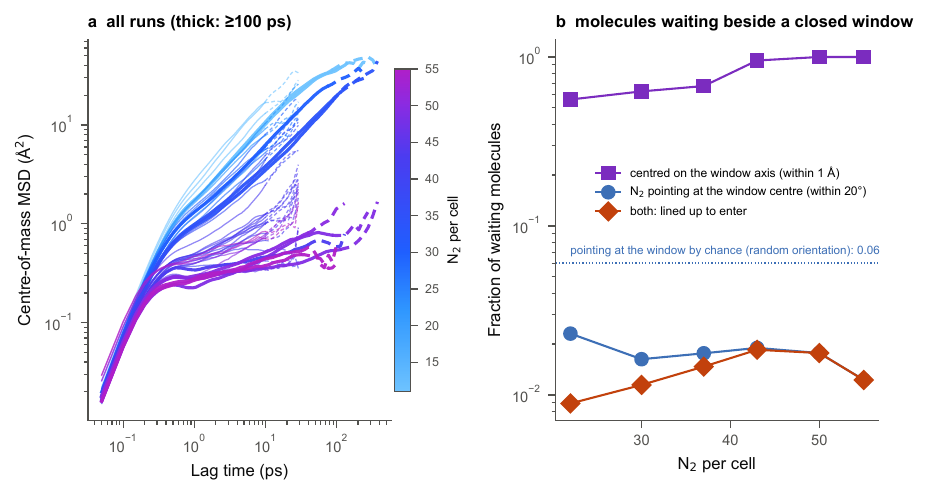}{Nitrogen mobility and the molecules waiting at the four-ring windows (uMOF-MH, 77~K). (a) Centre-of-mass MSD for every run, coloured by loading, to the longest lag each run allows (thick, runs of 100~ps or more; dashed, lags beyond a third of the run, where few independent time origins remain). (b) Of the molecules waiting beside a closed four-ring window (centre 3.5--5.25~\AA\ from the window centre), the fraction centred on the window axis (within 1~\AA), the fraction whose N$_2$ axis points at the window centre (within 20$^\circ$), and the fraction doing both, i.e.\ lined up to enter; dotted, the fraction that would point at the window by chance for random orientation.}{SI-fig:dynamics}

Within 50~ps the five molecules nearest each side of a window are still the nearest five 85--100\% of the time, so the crowd that holds a threaded molecule cannot rearrange to release it (Figure~\ref{SI-fig:dynamics}a).

\subsection{A spontaneous threading event}\label{SI-si:event}
\emph{Search.} Eight unbiased uMOF-MH runs at 43 and 50 \Ntwo\ and 200 or 250~K started from relaxed cells with every four-ring window empty and no linker turned (deuterium masses, cell fixed). Each run had 20~ps of Langevin dynamics and then a Nos\'e--Hoover chain, whose first 10~ps had to pass numerical checks on the temperature and on the drift of the conserved quantity; all eight passed. Window occupancy was logged every 0.05~ps.

\emph{The event.} In the 50 \Ntwo\ run at 200~K, one molecule entered an empty window 36.35~ps into the run and was still there at the end of the 456~ps analysed, apart from one excursion of 0.05~ps beyond the occupancy cutoff ( Figure~\ref{SI-fig:eventchain}; main-text Figure~5 and Movie~S1). Before entry it waited 3.5--4.6~\AA\ from the window centre, lying across the window axis. The two linkers whose methyl groups face it then swung to about 40$^\circ$ while the molecule turned end-on and moved from 4 to 0~\AA\ along the axis within about 1~ps; the other two linkers followed within a few tenths of a picosecond, and all four stayed at 29--38$^\circ$. The methyl groups alternate in direction round every four-ring, two pointing into each cage, as the $\bar{4}$ axis through the ring requires, so each side of a window has its own pair of gatekeepers.

\emph{History of the cell.} The cell is an empty-framework cell (50 \Ntwo\ packed into the empty framework, 30~ps of MD at 77~K, full relaxation). The window stayed empty and its ring unturned at every stage before the event, with the entering molecule 3.5--4.6~\AA\ from the window centre throughout.

\emph{Limits.} Both threading events (this one and that of Section~\ref{SI-si:event2}) come from one run at 200~K, in a single periodic cell of fixed size, with one potential. They show that the pathway exists without bias and ends in the threaded state; they do not give the rate at 77~K.

\sifigure{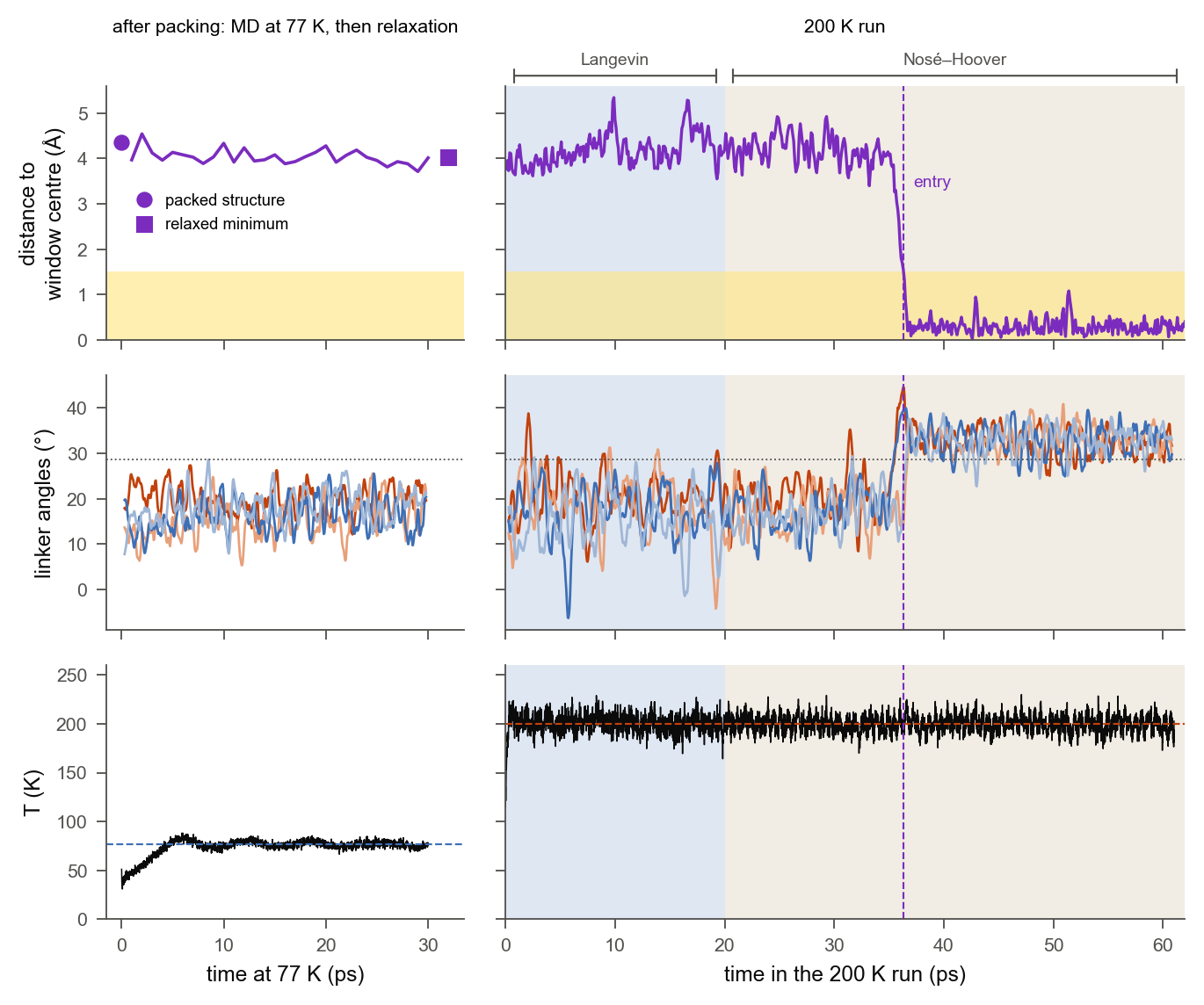}{History of the cell in which a molecule threaded a window. Left: 30~ps of MD at 77~K after packing (circle, packed structure; square, relaxed minimum). Right: the 200~K run, Langevin dynamics for 0--20~ps (blue shading) and a Nos\'e--Hoover chain from 20~ps to the end (beige); dashed line, entry. Top, distance of the entering molecule from the window centre, from the atomic frames every 0.1~ps (yellow, inside the window); middle, the four linker angles of that window (0.5~ps running mean; dotted, turned threshold); bottom, temperature.}{SI-fig:eventchain}

\subsection{A second threading event}\label{SI-si:event2}
Continuing the same 200~K run without any reset, a second molecule entered a different four-ring at 268.75~ps, from the side opposite to that from which the first entered its window. It had waited about 4~\AA\ from the window centre, entered within about 1~ps and was still there at 456~ps, apart from two excursions of 0.2~ps beyond the occupancy cutoff, while the first molecule stayed in its window. The mean linker angle of this ring, 20--25$^\circ$ over the whole logged run before the entry (23$^\circ$ over 262--268~ps), rose to 35$^\circ$ (Figure~\ref{SI-fig:event2}). The side of each methyl group was found from the atomic frames (two methyl carbons about 1.6~\AA\ on each side of the window plane, on opposite edges of the ring). As in the first event, the two linkers facing the molecule open first and the far pair follows as it enters (Figure~\ref{SI-fig:event2}).

\sifigure[.95]{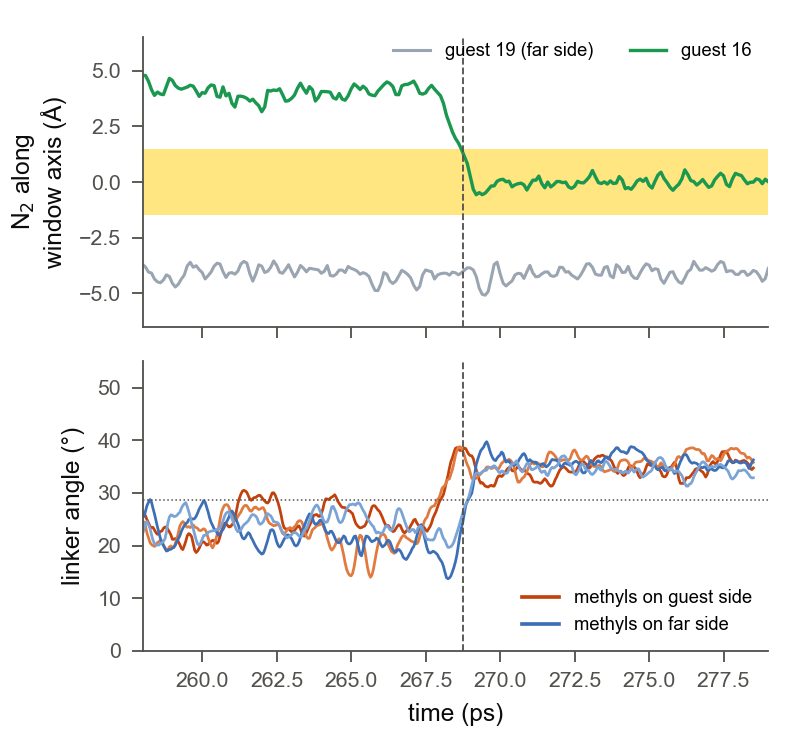}{Second threading event (uMOF-MH, 50 \Ntwo, 200~K, same run as main-text Figure~5). Top, position along the window axis of the threading molecule (green) and of the molecule waiting on the far side of the same window (grey), from the atomic frames every 100~fs; yellow, inside the window; bottom, the four linker angles (1~ps running mean; red and orange, methyl groups facing the molecule; blue, far side; dotted, turned threshold). Dashed line, entry at 268.75~ps.}{SI-fig:event2}

\subsection{Free energy of threading}\label{SI-si:umbrella}
\emph{Method.} The coordinate $s$ is the position of the centre of mass of one \Ntwo\ molecule along the normal to a four-ring window, recomputed at every step from the window's four Zn atoms, with $s=0$ at the window centre and $s<0$ on the side from which the molecule approaches. Each sampling window adds a harmonic bias $\tfrac{1}{2}k(s-s_0)^2$ with $k=0.15$~eV~\AA$^{-2}$, plus a flat-bottomed wall ($k=0.2$~eV~\AA$^{-2}$) that acts only if the molecule strays more than 2.5~\AA\ from the window axis. All runs used uMOF-MH with deuterium masses, the cell of the threading run, 200~K and a Nos\'e--Hoover chain.

The profiles of main-text Figure~5e use the threading cell itself (50 \Ntwo) and the same cell with two molecules added away from the four-rings (52 \Ntwo). At each loading, windows were started both from a frame of the unbiased run with the molecule waiting beside the window and from one with it threaded, so that the result does not depend on the starting side: 27 windows per loading from $-4.5$ to $+2.5$~\AA, 54 in all. Every window ran for 5~ps of equilibration and 50~ps of sampling; 44 of the 54 windows have so far been continued exactly to 150~ps of sampling; the remaining ten, eight of them beyond the window, still have 50~ps. The windows were combined with WHAM on 0.1~\AA\ bins, keeping bins with at least 250 samples (125 in each half), and the bands are $\pm1$ standard deviation from a block bootstrap (each window's trajectory cut into 10~ps blocks, about three times the correlation time of $s$, resampled with replacement within every window, 40 replicates). Beyond the outer barrier the free energy falls steeply and the window centred at $+2.5$~\AA\ slides to about 3.4~\AA, so 2.6--3.0~\AA\ is the least well sampled part of the profile.

\emph{Profiles.} Relative to the threaded molecule, the waiting molecule lies at $-61\pm3$~meV at 50 \Ntwo\ and $-14\pm4$~meV at 52 \Ntwo; the barrier back towards the waiting side is $131\pm3$ and $164\pm2$~meV, and onwards into the far cage $140\pm9$ and $164\pm13$~meV. Extending 44 windows threefold changed the barriers by at most 4~meV and the waiting state by up to 16~meV (52 \Ntwo). The threaded molecule therefore sits in a minimum guarded on both sides at both loadings, and the two added molecules narrow its free-energy cost by about 45~meV.

\emph{Energy and entropy.} The unbiased MLIP potential energy was recomputed on the saved frames every 100~fs and averaged with the umbrella weights (Figure~\ref{SI-fig:energyentropy}). Threaded minus waiting, $\Delta\langle U\rangle=-18\pm18$~meV at 50 \Ntwo\ and $-72\pm10$~meV at 52 \Ntwo, and $-T\Delta S=+77\pm24$ and $+84\pm9$~meV. The energy of the whole cell fluctuates by about 0.4~eV and drifts slowly as the crowd rearranges, so only basin-to-basin differences are well determined. Extrapolating $\Delta F=\Delta U-T\Delta S$ linearly in $T$ gives $-40$~meV at 52 \Ntwo\ and $+12$~meV at 50 \Ntwo\ at 77~K; this is a trend, not a prediction.

\emph{Orientation.} The angle $\theta$ between the \Ntwo\ axis and the window normal, reweighted in the same way (Figure~\ref{SI-fig:orientation}), shows the molecule lying across the window in the waiting basin (mean $|\cos\theta|=0.40$, against 0.5 for random orientation) and locked end-on from about $-2.5$~\AA\ through the window ($|\cos\theta|\approx0.93$). With $S_{\rm or}/k=-\int f\ln f\,\mathrm{d}x$ for the density $f$ of $x=|\cos\theta|$ (axial symmetry about the normal assumed), the orientational part of the entropy cost of threading is $-T\Delta S_{\rm or}=+29$~meV at 50 \Ntwo\ and $+28$~meV at 52 \Ntwo\ (200~K), about 40 and 30\% of the total $-T\Delta S$; the rest is the loss of lateral freedom in the window and changes in the surrounding crowd.

\sifigure[.95]{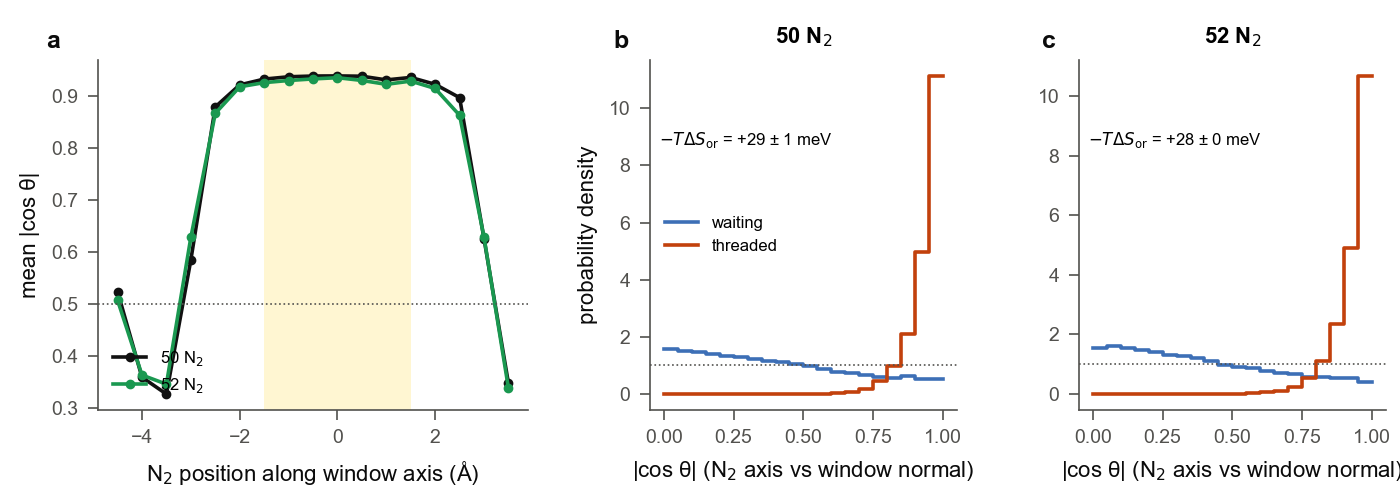}{Orientation of the threading molecule (umbrella sampling, 200~K, reweighted). (a) Mean $|\cos\theta|$ between the \Ntwo\ axis and the window normal along the window axis (dotted, random orientation; yellow, inside the window). (b, c) Distribution of $|\cos\theta|$ in the waiting (blue) and threaded (red) basins at 50 and 52 \Ntwo, with the orientational entropy cost of threading.}{SI-fig:orientation}

\emph{Gate.} The mean angle of each linker pair along the window (main-text Figure~5e, lower panel) shows the pair facing the waiting molecule opening on the approach, both pairs open while the molecule is in the window, and the far pair opening as it leaves.

\emph{Limits.} The cells are closed, so molecules cannot arrive from the gas and refilling is imposed rather than sampled. Differences of a few tens of meV are below the accuracy of current MLIPs, and all profiles use one potential.

\sifigure[.95]{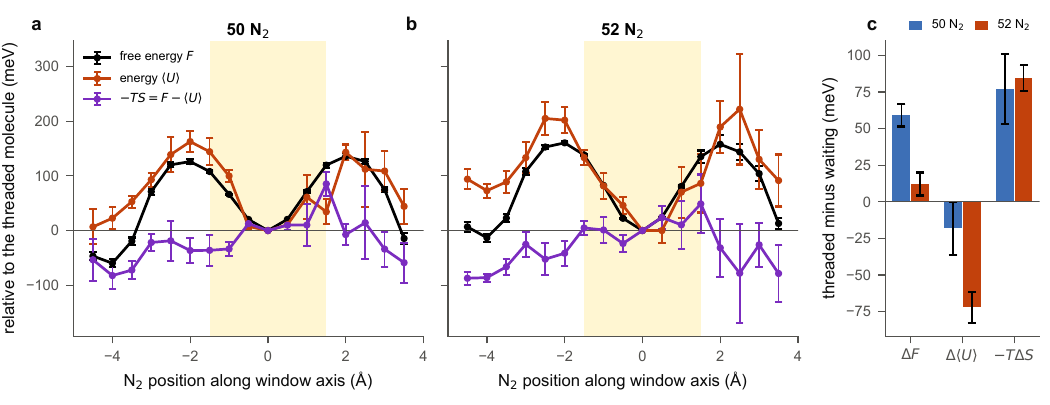}{Energy and entropy of threading at 200~K (umbrella sampling started from the waiting and the threaded frames, uMOF-MH; windows extended to 155~ps where available). The unbiased MLIP potential energy was recomputed on the saved frames every 100~fs and reweighted to the unbiased ensemble with the umbrella weights. (a, b) Free energy $F$, mean energy $\langle U\rangle$ and $-TS=F-\langle U\rangle$ along the whole window axis ($s=0$ at the window centre; yellow, inside the window) at 50 and 52 \Ntwo, relative to the threaded molecule; error bars, standard error over four consecutive quarters of each run. (c) Threaded minus waiting basin ($|s|<0.6$~\AA\ against $-4.6<s<-3.4$~\AA): $\Delta F$, $\Delta\langle U\rangle$ and $-T\Delta S$, with standard errors over the four quarters.}{SI-fig:energyentropy}

\subsection{Pore regions and passages between cages}\label{SI-si:regions}
\emph{Six-ring windows and passages between cages.} In the unbiased 200 and 250~K runs at 43 and 50 \Ntwo, five to seven of the eight six-ring windows of a cell hold a molecule at any moment, but the median visit lasts 0.5--0.8~ps and only 3\% (200~K) to 7\% (250~K) of visits end in the neighbouring cage. Every one of the 49 passages at 200~K and 98 at 250~K was made by a wall molecule, and only one of the passing molecules had visited the core in the preceding 20~ps (Figure~\ref{SI-fig:sixorigin}): the molecules that cross are long-standing wall molecules.

\sifigure[.85]{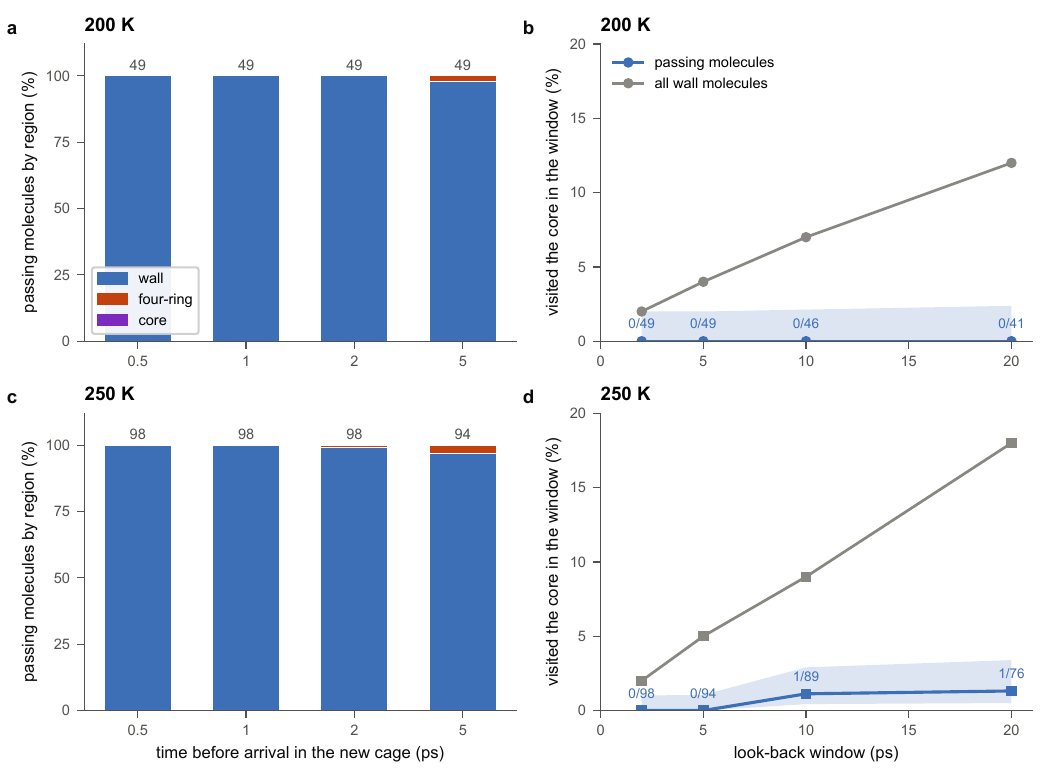}{Origin of the molecules that pass between cages through a six-ring, 200~K (a, b) and 250~K (c, d). (a, c) Region of each passing molecule a given time before it arrives in the new cage. (b, d) Fraction that visited the core in the preceding window (blue), against wall molecules in general (grey).}{SI-fig:sixorigin}

\emph{Regions (main-text Figure~5d).} Molecules of the threading cell are assigned, frame by frame, to three regions: the pocket in front of a four-ring (centre within 5.3~\AA\ of a four-ring centre, outside the window itself), the core (otherwise at least 4.75~\AA\ from every framework heavy atom) and the wall (the rest, including the six-ring windows). Both cutoffs lie at minima of the corresponding distance distributions. The threaded molecule is excluded. The density maps average 0.1--143~ps on a 0.25~\AA\ grid, in 2~\AA\ slices through the cage centre (the equatorial plane, and the $(1\bar{1}0)$ plane containing four six-ring windows). For the mean-squared displacement each molecule is assigned to its region at the time origin (50--250~ps, before the second threading event; origins every 0.5~ps; 68\% bootstrap bands over molecules). Fits over 4--10~ps give $9.3\times10^{-6}$ (core), $2.4\times10^{-6}$ (wall) and $1.8\times10^{-6}$~cm$^2$~s$^{-1}$ (four-ring pockets); the displacement grows as about $t^{0.6}$, so these describe local mobility rather than long-range diffusion.

\section*{Movie S1}
The first threading event (uMOF-MH, 50 \Ntwo\ per cell, 200~K, unbiased MD, 20--48~ps of the run), as in main-text Figure~5a--c; provided as the ancillary file \texttt{Movie\_S1.mp4}.

\section{Data and code}
Figure scripts, analysis scripts, input hashes and per-figure provenance records accompany the manuscript; the mode analysis uses read-only exports of the $\Gamma$-point eigenpairs and structures with recorded hashes. The uMOF-MH model file used throughout has the SHA-256 checksum \texttt{\small af6491a88c25740e5641633a54db64f9\allowbreak c8cfae320b8983be3a52c4958b29bcab}, so that the exact model can be identified. These local records are not yet a public deposition; experimental reduction records and source data will be released with the final version.

\renewcommand{\refname}{References for the Supporting Information}

\end{document}